\newcommand{\eg}{{e.g.,}\xspace}
\newcommand{\ie}{{\it i.e.,}\xspace}

\newcounter{NumTakeaways}
\stepcounter{NumTakeaways}

\newcommand{\numprofilesWeb}{{14}\xspace}

\newcommand{\numprofiles}{{24}\xspace}
\newcommand{\numphones}{{10}\xspace}

\newcommand{\amigo}{{\texttt{AmiGo}}\xspace}

\newcommand{\testbed}{{\texttt{AmiGo}}\xspace}

\newcommand{\redmi}{{Redmi-Go}\xspace}

\documentclass[acmsmall,nonacm]{acmart}

\setcopyright{none}
\usepackage{gensymb}
\usepackage{booktabs}  
\usepackage{pbox}
\usepackage{subfig} 
\usepackage{epsfig,endnotes}
\usepackage{epstopdf}
\usepackage{times}
\usepackage{color}
\usepackage{colortbl} 
\usepackage{graphicx}
\usepackage{xspace}
\usepackage{enumitem}
\usepackage{multirow}
\usepackage{tikz}
\usepackage[ruled,vlined,linesnumbered]{algorithm2e}

\AtBeginDocument{%
  }

\renewcommand\footnotetextcopyrightpermission[1]{} 
\begin{document}

\title{Unraveling the Airalo Ecosystem}

\author{HyunSeok (Daniel) Jang}
\email{hsj276@nyu.edu}
\affiliation{%
  \institution{NYU Abu Dhabi}
  \city{Abu Dhabi}
  \country{UAE}
}

\author{Matteo Varvello}
\email{matteo.varvello@nokia.com}
\affiliation{%
  \institution{Nokia Bell Labs}
  \city{New Jersey}
  \country{USA}
}

\author{Andra Lutu}
\email{andra.lutu@telefonica.com}
\affiliation{%
  \institution{Telefonica Research}
  \city{Madrid}
  \country{Spain}
}

\author{Yasir Zaki}
\email{yz48@nyu.edu}
\affiliation{%
  \institution{NYU Abu Dhabi}
  \city{Abu Dhabi}
  \country{UAE}
}
\renewcommand{\shortauthors}{Jang et al.}

\begin{abstract}
In recent years, we have witnessed myriad flavours of Mobile Network Aggregators (MNAs) which exploit the coverage footprint of a handful of base operators to provide global mobile connectivity. Under the MNA model, emerging operators reap the benefits of network softwarization and virtualization, including eSIM technology or control/data-plane separation. This paper investigates an emergent MNA type -- a \textit{thick} MNA -- that relies on multiple (core) base operators from different economies to provision eSIM profiles, while employing gateway functions to the public internet located outside the respective base operators' home country.  Specifically, our work is the first to capture the intricacies of Airalo -- a thick MNA that operates in 219 countries. 
Unlike other MNAs that our community scrutinized, we show that Airalo often decouples the geographical location of the public internet gateway from the native country of the base operator via IPX Hub Breakout (IHBO). To map Airalo's underlying infrastructure, we ran web-based measurements that \numprofilesWeb volunteers performed while traveling and using an Airalo eSIM on their personal devices. We further dive into Airalo's performance by running device-based measurements (speedtest, traceroute, video streaming, etc.) in \numphones countries with rooted Android devices. Finally, we examine Airalo's pricing by monitoring its marketplace. 
\end{abstract}




\maketitle
\section{Introduction}
\label{sec:intro}
Mobile Network Aggregators (MNAs) -- such as Google Fi, Twilio or Truphone -- have emerged in recent years as a disruptive phenomenon in the Telco sector. Their operational model upgrades the Mobile Virtual Network Operator (MVNO) approach~\cite{schmitt16pam, xiao19mobisys, zarinni14imc} to leverage existing infrastructure from multiple base operators in different countries, thus providing global mobile connectivity~\cite{alaca-marin2022mobileAggregators}. With the advent of embedded subscriber identity module (eSIM) technology -- integrated seamlessly into devices like smartphones and wearables -- some MNAs now enable travellers to avoid inconvenient (buying local SIM cards in stores) and expensive (roaming bill shock) connectivity while abroad. 

The combination of MNA and eSIM technology with the recent advances in network virtualization~\cite{jain22L25GC, luo21cellbricks, larrea23coreKube} enables a new era in global mobile connectivity, characterized by varying degrees of operational complexity. 
Until now, we have seen (commercial) evidence of MNAs that rely on the core networks of base operators (light MNAs) or run their own (full MNAs)~\cite{alaca-marin2022mobileAggregators, yuan18mobicom} to provision their eSIM profiles.
Both models gain access to (visited) radio access networks globally via interconnection through roaming hubs~\cite{lutu2020firstLook}.

In this paper, we provide the first empirical evidence of a new breed of MNAs -- \textit{thick MNA} -- that only run specific core network functions (\eg the gateway function to the public internet access), while still relying on the cellular ecosystem, similar to earlier models. 
Airalo~\cite{airalo} is a popular thick MNA that has gained more than 5 million customers since its inception in 2019.  Airalo decouples the internet gateway location from both the base and the visited operators' infrastructure. 
Despite its popularity, anecdotal reports on Airalo's quality of service have been largely mixed~\cite{redditAiralo}.

Motivated by the above, our paper investigates thick MNA operations by dissecting Airalo from \textit{infrastructural}, \textit{performance}, and \textit{economic} perspectives. 
We acquired Airalo eSIM plans for 14 countries, and had volunteers use them while traveling, performing high-level measurements via JavaScript on our website~\cite{ourPage}. To compare with local physical SIMs, we also had volunteers carry a rooted Android device executing automated tests in 10 additional countries. This allowed us to measure 11\% of Airalo’s global footprint, covering 24 of the 219 served countries.
Finally, we examined Airalo’s economic background by monitoring eSIM offers (cost, provider, and data limit) on eSIMDB~\cite{esimdb} over four months (from February to May 2024) to identify pricing trends and potential discrimination. Additionally, with the help of 10 volunteers, we manually collected data on local SIM offerings.





Our analysis reveals that six (base) operators provision Airalo eSIMs operating via roaming in 21 of the 24 countries we measure (for the other three countries, the Airalo eSIMs are native profiles issued by a local operator). This extensive roaming network enables Airalo to reach global coverage without completing direct (and usually long-drawn-out) agreements with local mobile operators in each country. 
However, roaming enforces packets to route through dedicated Packet Data Network Gateways (PGWs) that may be geographically distant from the user (usually in the core network of the base operator~\cite{mandalari2018Roaming, fezeu2024roaming5GinEU}). 
For the roaming eSIMs, we infer their topology by mapping their public IP to the associated Autonomous System number (ASN). 
We find a prevalent use of both Home Routing (HR) and IPX Hub Breakout (IHBO) as the underlying roaming topology, as well as the lack of Local Breakouts (LBO). 

Next, we summarize the main takeaways from our analysis.

\vspace{0.05in}
\noindent
\textbf{Limited PGW Selection.} 
To the best of our knowledge, this is the first study to show evidence of IHBO used by a commercial operator. 
In theory, IHBO can enable dynamic routing of roaming traffic, prioritizing PGWs closer to (but not belonging to) the \textit{visited Mobile Network Operator} (v-MNO)~\cite{US20140169286A1, lutu2021insights, alaca-marin2022mobileAggregators}. 
In practice, we find that the PGW locations are restricted via pre-configured agreements among MNOs, IPX-Ps and PGW providers, limiting the possible performance benefits of using IHBO
(Section~\ref{sec:res:network}).
Most Airalo eSIMs rely on a single, fixed PGW provider, indicating a static pre-arrangement of PGW selection.

\vspace{0.05in}
\noindent
\textbf{Latency Degradation.} Popular providers like Google and Facebook place edge nodes close to PGWs operated by major IPX-Ps and MNOs. As such, we find that Airalo (roaming) eSIMs tend to have comparable \textit{public} distance -- \ie number of hops on the public internet after breaking out -- with respect to their physical SIM counterparts (Section~\ref{sec:res:network:path}). However, this is insufficient to offset the preceding \textit{private} distances between end user and PGWs, which are often due to the inefficient PGW selection discussed above. Approximately 14.5\% of the latency measurements done via roaming eSIMs provide \textit{less desirable} latencies (\ie exceeded 150 ms), while only 3\% of measurements done in physical SIM cards reached such high latency levels.
However, IHBO reduces the impact of home routing latency through the base operator's infrastructure. Compared to the native setup, IHBO inflates the latency by 64\%, an order of magnitude less than the 621\% increase of home routing, the standard setup for roaming~\cite{mandalari2018Roaming, fezeu2024roaming5GinEU}. 

\vspace{0.05in}
\noindent
\textbf{Negligible Impact on Data Speeds.} We find that the underlying roaming configuration for the Airalo eSIMs did not correlate with the end-user's network bandwidth (Section~\ref{sec:performance}). Specifically, IHBO did not lead to significant improvement over home routing in terms of download speed. Overall, however, we find considerable variation in the downlink among eSIMs from the same base MNNO (b-MNO), \eg spanning from 22~Mbps in Germany up to 112~Mbps in Spain, despite both being provisioned from \textit{Play Poland}. 
These findings highlight that network throughput for roaming eSIMs is largely contingent upon the policies of the v-MNO, rather than the specific roaming topology chosen.

\vspace{0.05in}
\noindent
\textbf{Consistent Pricing Strategy.} Airalo maintains a relatively consistent pricing strategy across different regions, with noticeable variations in cost between continents (Section~\ref{sec:economics}). Despite being one of the largest eSIM providers with 219 countries covered,  
Airalo's pricing ranks moderately high compared to its competitors (11th highest out of 54 providers, with median cost per GB of \$7.9). 


\section{Background and Related Work}
\label{sec:related}
This section provides some useful background for a clear understanding of the  paper. We also position our contributions in relation to relevant related work. 


\vspace{0.05in}
\noindent
\textbf{eSIM Technology.} Traditional mobile communication relies on \textit{physical} SIM cards, which store subscriber details like the International Mobile Subscriber Identity (IMSI), authentication keys, and network authorization data. Each SIM card is tied to a specific MNO. In contrast, embedded SIMs (eSIMs) are integrated directly into devices, capable of hosting multiple subscriber profiles, or ``eSIM profiles``, without requiring physical swapping. This capability is enabled through Remote SIM Provisioning (RSP)~\cite{GSMA2018_esim}. 
While eSIM technically refers to the SIM hardware within the device, this paper will use the term ``eSIM'' to refer to an ``eSIM profile'' for brevity.

RSP creates a marketplace of eSIM providers, acting as intermediaries between users and eSIM owners like MNOs. 
eSIM providers offer user-friendly web and mobile applications for purchasing customized eSIM plans based on service type, duration, and geographical coverage.
The eSIM marketplace includes, at the time of writing, 54 providers~\cite{esimdb}.

\begin{figure}
    \includegraphics[width=0.85\columnwidth]{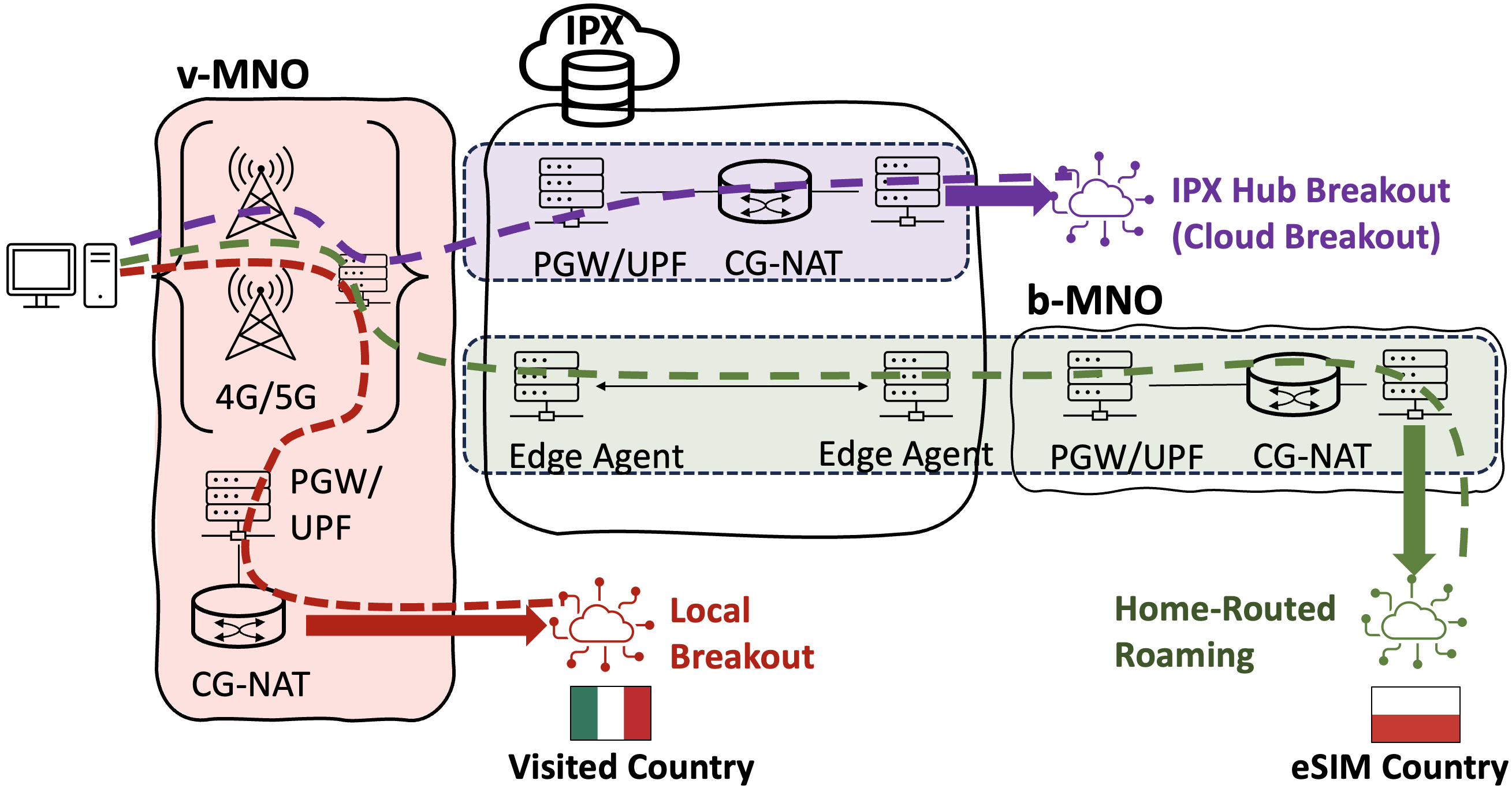}
    \vspace{-2mm}
    \caption{Roaming configuration for the data path of an eSIM provisioned by a b-MNO in Poland (green), and operating in Italy (red) via a local v-MNO: home-routed roaming routes the traffic through the home country, local breakout routes the traffic through the visited country, and IPX hub breakout routes the traffic via a third-party location offered by the IPX Provider (or a cloud provider).}
    
    \label{fig:mobile_data_architecture}
\end{figure}

\vspace{0.05in}
\noindent
\textbf{Roaming.}  MNOs support international mobility of their users via \textit{roaming}.  
In the context of mobile communication, the terms \textit{native} and \textit{roaming} user distinguish between the network to which an end user connects. 
A native user is connected to the b-MNO's network, the infrastructure of the MNO that issued the SIM. 
Conversely, a roaming user is outside the coverage area of the b-MNO. To access mobile services in a foreign visited country, the device connects to a v-MNO that has a roaming agreement with the b-MNO.

MNOs typically rely on the IP Packet Exchange (IPX) network ~\cite{GSMA_IPX_whitepaper1, GSMA_IPX_whitepaper2, HuaweiLTERoaming2024} to facilitate world-wide roaming services. 
This consists of a small set of IPX Providers (IPX-P) that peer with each other over a private IP backbone, forming a tightly meshed network isolated from the public internet~\cite{lutu2021insights, lutu2020firstLook}. 
By contracting an IPX-P, an MNO 
can expand their service footprint globally through a single point of contact.  
The connections between MNOs can be further configured in accordance to their roaming agreements, with the IPX-P enabling policies to control roaming services.


Figure~\ref{fig:mobile_data_architecture} shows the architectural differences of the three main roaming architectures~\cite{US20140169286A1, mandalari2018Roaming, mandalari2021measuring}. \textit{Home-Routed Roaming} (HR) involves assigning the public IP address of a roaming user by the b-MNO. All inbound and outbound traffic flows through a GTP tunnel between a Serving Gateway (SGW) within the v-MNO and a PGW of the b-MNO. While this allows access to IP-based services provided by the b-MNO, it comes with higher latency due to the GTP tunnel. \textit{Local Breakout} (LBO) allows the v-MNO to assign the public IP address of a roaming user, enabling direct internet access without traversing the IPX network to reach the b-MNO. However, LBO may limit access to b-MNO-specific services and prevent service control and charging from the b-MNO.



\textit{IPX Hub Breakout} (IHBO) uses PGWs hosted by a third-party within the IPX network to assign public IP addresses to roaming users. A GTP tunnel links the v-MNO's SGW to a selected PGW based on factors like geography, latency, and business agreements. The IPX network, trusted by the b-MNO, assigns an IP address recognized by the latter to support operator-specific services for the roaming user.

\begin{figure}[!t]
    \includegraphics[width=0.5\columnwidth]{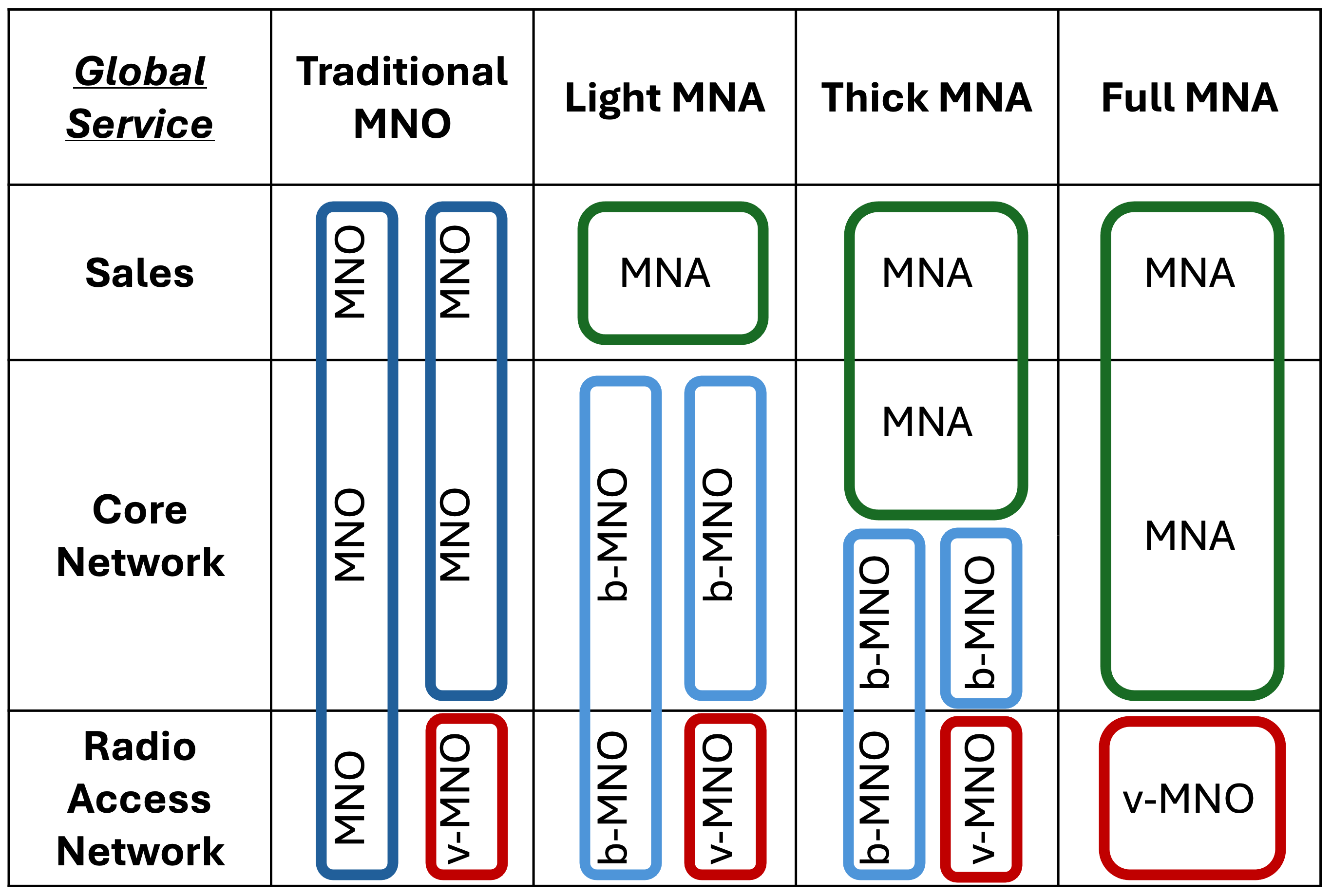}
    \caption{\small MNAs run a limited part of the network (the \textit{light} -- only sales $<<$  the \textit{thick} -- limited core function $<<$ the \textit{full} -- all the core), and provide global service by exploiting the roaming agreements of several b-MNOs;  our paper is the first to investigate a thick MNA. }
    \label{fig:mna_models}
\end{figure}

\vspace{0.05in}
\noindent
\textbf{Mobile Network Aggregators (MNAs).} eSIM providers like Airalo function as MNAs, \ie they partner with few b-MNOs to sponsor the eSIMs that travelers can use world-wide, either in roaming or native configuration.  Figure~\ref{fig:mna_models} captures the differences between MNA flavors. 
The MNAs run a limited part of the network: the \textit{light} -- only sales, the \textit{full} -- sales and full core deployment, and the \textit{thick} -- sales and a limited part of the core. Their global service relies on aggregating the international footprint of several b-MNOs (\ie for the light/thick model), or accessing directly one or more IPX-P for roaming hub services (\ie for the full model). 

Airalo decouples the b-MNO that provisions the eSIM from the traditional notion of the \textit{home} MNO of the user: we show that the b-MNO may be from a different country than the home or the visited countries of the end-user, breaking apart from other models of MNAs. For example, a user from the US travels to Italy, where their Airalo eSIM uses a profile issued from a Polish operator (b-MNO) that connects locally to an Italian provider (v-MNO).

\vspace{0.05in}
\noindent
\textbf{Comparison With Related Works.} Several studies have shown that roaming users suffer from penalties in network performance and Quality of Experience (QoE). In~\cite{mandalari2021measuring, mandalari2018Roaming, Michclinakis2018cloudServices, fezeu2024roaming5GinEU}, physical SIM cards from popular MNOs in Europe were found to rely on HR for roaming, causing additional latency with respect to the geographical distance between the roaming user and the b-MNO. Our work differs from these papers in two ways. First, we focus on a thick MNA operating at a global scale (Airalo). Second, we discover that significant portion of eSIM plans in Airalo depends on international roaming, with a mix of HR and IHBO configuration, which had not been previously observed. Accordingly, we offer the first comprehensive comparison of these two roaming architectures with respect to network performance. 

Our work is closely related to~\cite{alaca-marin2022mobileAggregators}, which analyzes the roaming performance of three MNAs: Google Fi -- a light MNA via eSIMs, Truphone -- a full MNA via eSIMs, and Twilio -- a full MNA via physical SIMs. The paper shows that, back in 2022, all providers relied on HR for roaming. However, the measurements were limited to roaming in two EU countries (Spain and Norway) using an international roaming plan from each MNA. In comparison, our work has larger geographical coverage -- spanning 24 countries across 4 continents -- and measures the performance of eSIMs tied to specific regions, which may or may not involve roaming. We also note that Airalo targets a different user demographic from Google Fi, which is limited to mobile customers in the US, or from Twilio, which is limited to IoT devices. Indeed, our paper is the first to study the reality of a thick MNA, which aggregates services from MNOs globally and exploits their roaming agreements, but also involves the IPX network. 

\section{Methodology}
\label{sec:meth}
This section describes the methodology we have devised to shed light on Airalo, to the best of our knowledge, the most popular implementation of a thick MNA. Our methodology captures three fundamental aspects: 1) infrastructure, 2) performance, and 3) economics. We have devised three distinct measurement campaigns, which we refer to as: web-based, device-based, and crawler-based.  

\subsection{Web-Based Campaign}
We ran the web-based measurement campaign from March 10th to 22nd, 2024, to investigate the infrastructure and some high-level performance of Airalo. This campaign targeted \numprofilesWeb countries (9\% of Airalo's current offering), selected based on the opportunity to have volunteers visiting such countries and use a complimentary eSIM.  Accordingly, we acquire \numprofilesWeb Airalo eSIM plans for which we identify the v-MNO as the network operator displayed in the phone (while travelling), and its b-MNO as the MCC-MNC codes from the Access Point Name in the device settings. 
While travelling, volunteers access a webpage~\cite{ourPage} we developed; the webpage first prompts to upload a screenshot of network settings, which is analyzed with chatGPT vision~\cite{gptVision} to verify that the device is using the provided eSIM (and not Wi-Fi). If successful, the next step consists in testing the DNS configuration via Nextdns~\cite{nextdns}.  Note that we ensure that the volunteers have no custom on-device DNS settings. 


Finally, we perform a speedtest by loading \url{https://fast.com} in an iframe, as it is the only popular speedtest provider that permits iframe embedding. We then ask the volunteers to upload a screenshot of test results upon completion; we again use chatGPT vision to analyze the screenshot, verify correctness, and extract data like download speed and latency.  We use the public IP associated with an eSIM -- derived both via  \url{fast.com} and logged by our server -- to characterize the roaming architecture by matching its Autonomous System Number (ASN) against the b-MNO's (HR), the v-MNO (LBO), or a third party such as an IPX-P (IHBO). 

\subsection{Device-Based Campaign}
The device-based measurement campaign was conducted from December 2023 to April 2024, providing 10 volunteers with rooted Android phones equipped with 
both an eSIM (purchased from Airalo) and a local SIM from the same v-MNO used by the eSIM. These phones, supposed to be carried and not used by our volunteers, are set up to run a plethora of measurements while switching between physical SIM and eSIM. The main goal of this campaign is to characterize the performance of Airalo, especially when compared with local SIM providers. 


To deploy the above measurements, we extend the (open-source) \testbed code~\cite{amigoPaper}, which provides  a control server to remotely manage mobile measurement endpoints (MEs). The control server offers restful APIs which the MEs call to: 1) report their current status (\eg battery level and connectivity), and 2) retrieve instrumentation code. The MEs are rooted Android devices instrumented via \texttt{termux}~\cite{termux}. \amigo only supports the \redmi, which is unsuitable for our measurements as it lacks eSIM support. We identify the Samsung 21+ 5G as a suitable alternative, given its ease of rooting and eSIM compatibility. We subsequently extend \amigo to support this device, implementing new automation hooks which will also be open-sourced.

The devices 
are instrumented to periodically run 
various network experiments 
which we summarize in the following. The interested reader can refer to~\cite{amigoPaper} for a more detailed description of such experiments. At low level, we measure download/upload speed and latency. When needed, we use as endpoint Ookla servers, via speedtest~\cite{speedtest}, and popular content providers characterized by a large footprint like Cloudflare and Google. Going up the stack, we measure DNS performance and settings, as well as the performance of popular CDN providers.  We further develop a browser extension that injects JS code to enable YouTube's ``stats-for-nerds'', which provides detailed statistics such as video quality, download speed, and buffer occupancy. This extension is installed on the Kiwi browser~\cite{kiwi}, which supports Chrome extensions on mobile platforms; Kiwi is further automated to open a YouTube link to a video with a resolution of (at most) 4K~\cite{ourVideo}. Similar to the web-based campaign, the public IP address obtained from these experiments are used to determine the roaming architecture of the activated eSIM. Finally, we derive insights on network paths by running traceroutes to global service providers (\eg Google) using mtr~\cite{mtr}.

\begin{table}
\centering
\small

\setlength\tabcolsep{0.5pt}
{\renewcommand{\arraystretch}{1}
\begin{tabular*}{\linewidth}{@{\extracolsep{\fill}} c|c|c|c|c }
 \textbf{Visited Countries} & \textbf{b-MNO (Country)} & \textbf{PGW Provider(s) (ASN)} & \textbf{PGW Country(s)} &\textbf{Type}\\
 \hline
ARE, JPN, PAK, MYS, CHN&Singtel (SGP)& Singtel (AS45143)  & SGP &HR\\
 \hline
 GBR, DEU, GEO, ESP& Play (POL) & Packet Host (AS54825)    &NLD &IHBO\\
 & & OVH SAS (AS16276)& FRA &IHBO\\
 \hline
QAT, SAU, TUR, EGY & Telna Mobile (USA) &Packet Host (AS54825) & NLD&IHBO\\
 & & OVH SAS (AS16276)& FRA &IHBO\\
 \hline
 MDA, KEN, FIN, AZE & Telecom Italia S.p.A (ITA) & Wireless Logic (AS51320)&GBR&IHBO\\
 \hline
ITA, USA &  Orange S.A (FRA)& Webbing USA (AS393559)  & NLD, USA &IHBO\\
 \hline
FRA, UZB& Polkomtel Sp. z o.o. (POL)&   Packet Host (AS54825)  & USA&IHBO\\
 \hline
\end{tabular*}}

\vspace{0.05in}
\caption{\small ISO codes of visited countries (first column) with roaming eSIMs that have the same b-MNO (second column). The third column refers to their PGW Providers identified via our measurement endpoints' public IP addresses, with their country-level geolocation shown in the fourth column. The last column denotes the roaming architecture.}
\label{Tab:tab1}
\vspace{-0.1in}
\end{table}

\subsection{Crawler-Based Campaign}

The final subgoal of this paper is to characterize the economic aspect of Airalo, especially when compared to local SIM providers.  We develop a web crawler targeting EsimDB~\cite{esimdb} which  aggregate offers for eSIMs, capturing key variables like cost, provider, and data limits from a comprehensive set of eSIM providers spanning 244 regions. These regions include various geographical levels, including country (\eg UAE), subcontinent (\eg Gulf), and continent (\eg Asia). 

We conducted daily retrievals of eSIM offers over a four-month period from February to May 2024, identifying 54 unique eSIM providers. We further run the crawler at three different physical locations (Spain, New Jersey, and UAE) in April/May 2024 to investigate potential price discrimination tactics. Discovering local SIM offerings is instead more challenging since no global aggregator (like EsimDB) exists.  Accordingly, we resort to Web information and insights from the volunteers travelling to the countries of our experiments. 

\section{Airalo Tomography}
\label{sec:res:network}
This section analyzes data from both the web-based and device-based campaigns (from December 2023 to April 2024) with the goal of shedding some light on the underlying network infrastructure supporting Airalo’s service.
\subsection{Network Architecture}
\label{sec:res:network:arch}
We dissect Airalo's network architecture using measurements from \numprofiles nation-level eSIMs. 
We first examine the role of the v-MNO; when the v-MNO and the b-MNO are identical, the user can access mobile data as a ``native'' v-MNO user (non-roaming), hence we refer to these eSIMs as \textit{native} eSIMs. 
This applies to only three of the \numprofiles eSIMs from our two campaigns: LG U+ (South Korea), Ooredoo Maldives (Maldives), and dtac (Thailand). 
In these countries, Airalo relies on the b-MNO to issue the eSIMs, essentially acting as a light mobile virtual network operator~\cite{li2020understanding} ``renting'' the (RAN and core) infrastructure of the corresponding b-MNO.

\begin{figure*}
    \centering
    \includegraphics[width=\linewidth]{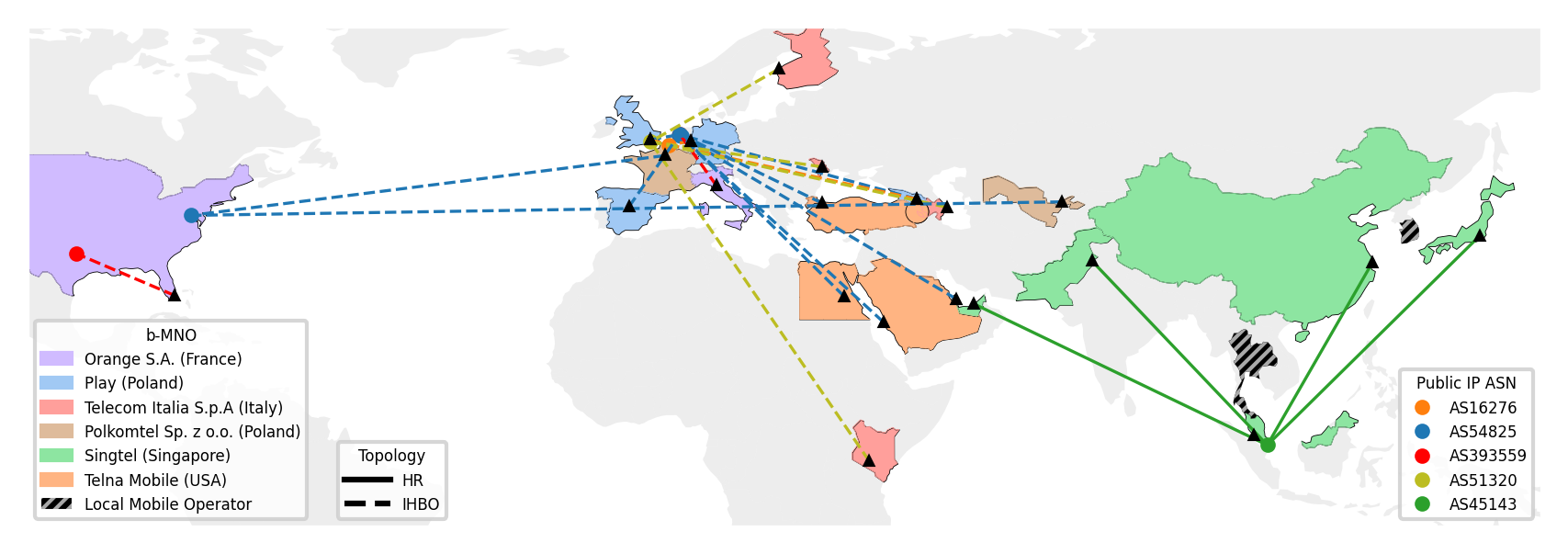}
    \vspace{-20pt}
    \caption{\small Mapping of end user location (triangle) and PGW location (circle) for 21 roaming eSIMs from Airalo. Each line visualizes the approximate distance between SGW and PGW for each eSIM, and the roaming architecture in use: solid line for HR, dashed line for IHBO. The color of PGW location and connected lines correspond to the PGW Provider (see legend in the lower right). Each country is colored based on the b-MNO (see legend in the lower left) associated with an eSIM.}
    \vspace{-2mm}
    \label{fig:map1}
\end{figure*}
For the remaining 21 eSIMs, the end user registers in the v-MNO as a roaming user from a b-MNO. Data roaming must be enabled for these eSIMs, hence we  refer to them as \textit{roaming} eSIMs. We find that six b-MNOs provision these 21 roaming eSIMs. Five eSIMs roam via Singtel, a Singapore-based MNO, using HR (\ie their public IPs belong to AS45143 from Singtel). The remaining 16 profiles employ IHBO, with data packets breaking out via PGWs provided by third-party infrastructure owners, such as the IPX-Ps. 

Table~\ref{Tab:tab1} summarizes the b-MNO, PGW provider(s), and PGW location(s) of the 21 roaming eSIMs. 
Each row corresponds to a group of visited countries where the eSIMs shared the same b-MNO and PGW providers. Note that for eSIMs provisioned by \textit{Play} and \textit{Telna Mobile}, the PGW provider iterated between 
Packet Host (AS54825) and OVH SAS (AS16276). 
PGW locations were inferred from geolocation of the public IP assigned to device using each eSIM. 
We verify this approach with traceroute analysis in section~\ref{sec:res:network:path}. 

Figure~\ref{fig:map1} visualizes the network infrastructure supporting Airalo’s service, given our visibility in eSIMs from \numprofiles visited countries. Each visited country is color-coded according to its b-MNO, \eg Italy and USA are colored in purple as their eSIMs both rely on Orange S.A (France) as b-MNO. A black triangle indicates the city where the eSIM was used by our volunteers, approximating the SGW location within the v-MNO. We mark PGW locations with circles, colored according to their ASNs. For example, there are 2 PGW locations for AS54825 (Packet Host, which supports IPX peering) -- one in Amsterdam (Netherlands) and one in Ashburn (Virginia, USA) -- which are both colored in blue. Each line connecting a black triangle to a PGW location illustrates the straight line distance between SGW and PGW for a given eSIM. This depicts the GTP tunnel traversed by packets before internet breakout, not to be confused with the data path from PGW provider's network to public servers as captured in traceroutes. 
We match a line color with the PGW location, and we use a solid line to indicate HR, and a dashed line for IHBO.

\subsection{IHBO Configurations}

When we detect a roaming eSIM, the network setup for the b-MNO is a mix between HR and IHBO. We did not detect any eSIM using LBO, likely due to a lack of trust among MNOs regarding roamer records and charges~\cite{lutu2020dice}. In the following, we present the main insights from dissecting Airalo's roaming eSIMs from a topology perspective. 


\vspace{0.05in}
\noindent \textbf{Suboptimal Choice of Roaming Configuration.} IHBO aims to optimize roaming traffic by directing packets to an IPX-P PGW located near the v-MNO. However, for 8 out of the 16 eSIMs we identified as IHBO, packets break out in IPX-P PGWs that are farther away from the end user location than the b-MNO country. We further observe that the PGW selection within an IPX-P is often not geographically optimal. 



\begin{figure*}
    \includegraphics[width = 1\linewidth]{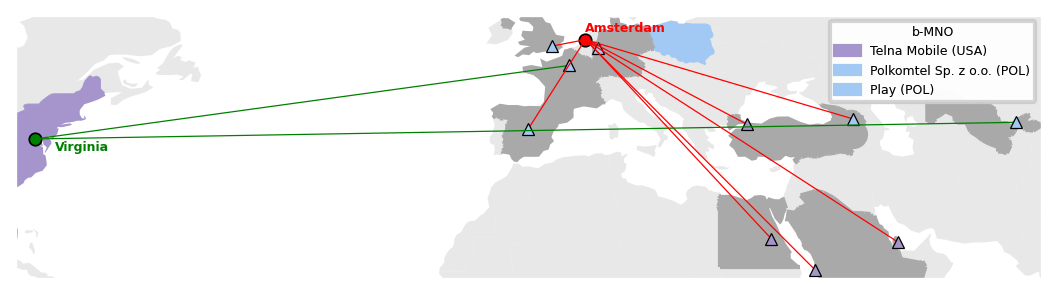}
    \vspace{-10pt}
    \caption{\small
    Mapping of end-user locations (triangles) and PGW locations (circles) for 10 eSIMs using Packet Host (AS54825) as PGW provider. End-user locations are colored by b-MNO country (see legend). PGW locations are also color-coded: red for Amsterdam, green for Virginia.    
    }
    \centering
    \vspace{-10pt}    
    \label{fig:map3}
\end{figure*}

Figure~\ref{fig:map3} shows end user and PGW location for 10 eSIMs, whose PGW provider was Packet Host (AS54825). Each end user location is marked with a triangle, colored according to the b-MNO country. We visualize the distance between SGW and PGW with one line per eSIM, colored based on PGW location. Notably, packets from France and Uzbekistan broke out in Virginia (US), despite the availability of closer PGWs operated by the same IPX-P in Amsterdam (Netherlands), which is used by eSIMs issued by Play and Telna Mobile. Given that the eSIMs for France and Uzbekistan were issued by Polkomtel Sp. z o.o., this suggests that the PGW location is decided based on the b-MNO. The figure further shows that  IHBO directs traffic from the Turkey eSIM to Amsterdam, further away from its b-MNO network in Poland.

\begin{figure}
\centering
    \begin{minipage}[t]{0.45\linewidth}
      \centering
      \includegraphics[width=1\textwidth]{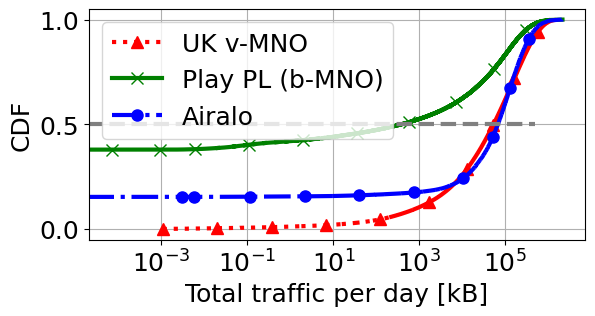}
      \vspace{-3mm}
      \label{fig:figure1} 
    \end{minipage}
    \hfill
    \hfill
    \begin{minipage}[t]{0.45\linewidth}
      \centering
      \includegraphics[width=1\textwidth]{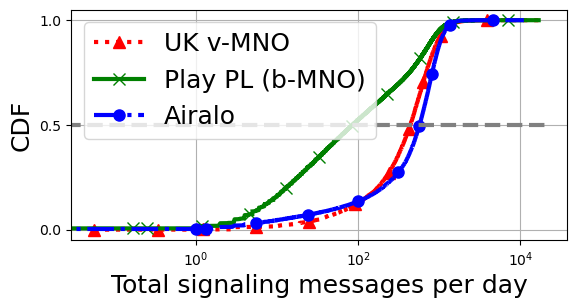}
      \vspace{-3mm}
      \vspace{-4mm}
      \label{fig:figure2}
    \end{minipage}
    \hfill
    \vspace{-5mm}
\caption{\small Data/signalling traffic comparison between inbound roamers from Airalo with those from Play (b-MNO). }
\label{fig:mno_visibility}
\end{figure}

\vspace{0.05in}
\noindent\textbf{Compromised v-MNO Visibility.} A v-MNO is unaware of the intricacy of the Airalo service, so it identifies Airalo users as inbound roamers from the b-MNO provisioning their eSIM. This obfuscates the Airalo users' demand patterns, and hinders network intelligence from the v-MNO side. 

To illustrate this challenge, we collaborate with a UK MNO and investigate the behavior of Airalo users. To identify Airalo users from the group of b-MNO inbound roamers, we deploy 10 eSIMs which rely on Play Poland as b-MNO, while using our cooperating MNO as a v-MNO in the UK. We then verify from the v-MNO core the IMSI range associated with these eSIMs, and extract a pattern matching potential IMSIs the b-MNO allocated to Airalo. 

Figure~\ref{fig:mno_visibility} compares the traffic consumption of the potential Airalo users in the UK -- identified via IMSI pattern matching -- connecting to our partner v-MNO, with the inbound roamers from Play Poland. As a reference, the figure also shows the traffic consumption of regular non-roaming users of the UK v-MNO. This analysis captured the user activity during April 2024. The figure shows that in terms of traffic consumption (both for data in Fig.~\ref{fig:mno_visibility}.a and signalling in Fig.~\ref{fig:mno_visibility}.b), Airalo is similar to the v-MNO's native users. Instead, Play Poland roamers show a different behavior, probably since they rely on multiple v-MNOs in the UK (not only the one we analyze).  Interestingly, the signaling traffic volume for inferred Airalo users is slightly higher than what v-MNO's native users generate (Fig.~\ref{fig:mno_visibility}.b), which is problematic for the v-MNO (signaling traffic in roaming is not charged) and might increase the device's energy consumption. This observation demonstrates how Airalo can add noise to v-MNO network intelligence.



\subsection{Path Analysis} 
\label{sec:res:network:path}
To further dissect the network infrastructure supporting Airalo's service, we rely on traceroute executed by the \numphones mobile devices from the device-based campaign. Each device was equipped with both physical SIMs and eSIMs, and measured two popular service providers (SPs) -- Facebook and Google -- which we chose for their global footprint.  We use \texttt{nslookup} to resolve the IP address associated with each hop identified by traceroute; for public IPs, we  obtain their geolocation and ASN using WHOIS information and ipinfo~\cite{ipinfo}. 

All observed paths begin with a variable number of hops associated with private IPs. 
Given the lack of visibility into GTP tunnels in end-to-end measurements~\cite{lutu2020firstLook}, these hops represent internal routing within the PGW provider's core network. The first hop is likely the PGW itself, followed by eventual forwarding to a Carrier-Grade Network Address Translation (CG-NAT), where packets are assigned globally routable IP addresses before internet breakout. 

\begin{figure*}
\centering
\begin{minipage}[t]{0.58\linewidth}
    \centering
    \includegraphics[width=1\textwidth]{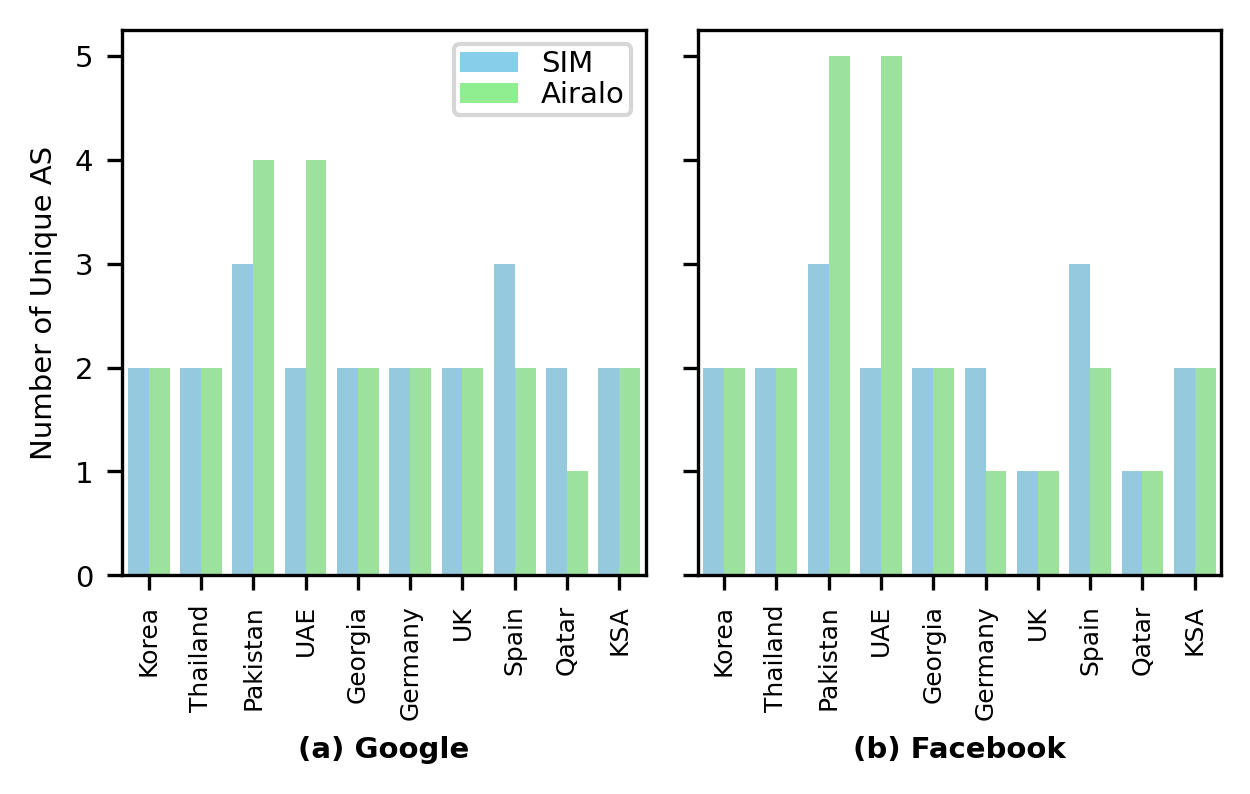}
    \vspace{-5mm}
    \caption{Median number of unique AS numbers observed in traceroutes to Google (left) and Facebook (right); SIM versus Airalo.}
    \label{fig:unique_asn}
\end{minipage}
\hfill
\begin{minipage}[t]{0.36\linewidth}
    \centering
    \includegraphics[width=1\textwidth]{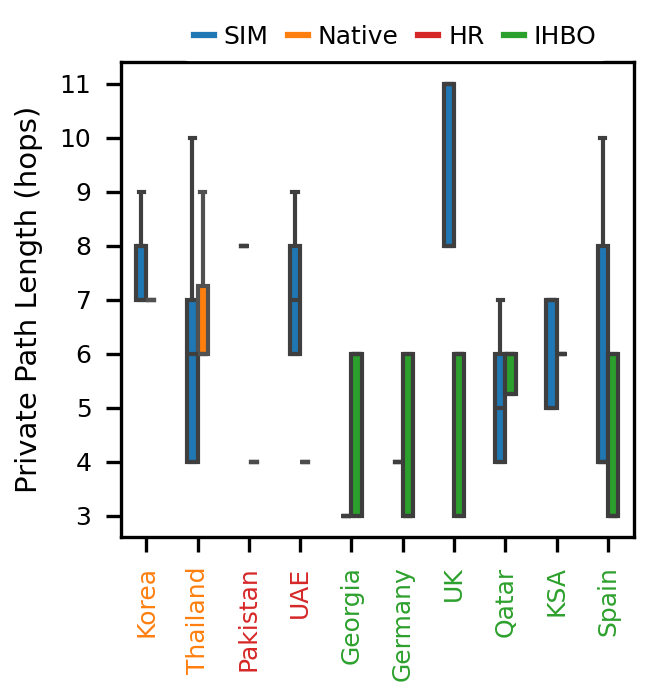}
    \vspace{-5mm}
    \caption{\small Private path length per country, analyzed from traceroutes to Google.}    
    \label{fig:private_hops}
\end{minipage}
\end{figure*}

To identify the PGW provider, we refer to the ASN of the first public IP address in each traceroute. We verify this methodology by confirming the ASN matches that of the device's public IP address, obtained from Ookla Speedtest performed shortly before each traceroute~\cite{amigoPaper}. This step renders impossible scenarios where the PGW and CG-NAT are hosted in different networks, which could lead to misclassification of network configurations (\eg b-MNO's PGW forwards packets to a CG-NAT in another AS, resulting in misclassifying HR traffic as IHBO).  In fact, Figure~\ref{fig:unique_asn} shows that most traceroutes reveal just two unique ASNs: one associated with the PGW provider, and the other associated with the SP (Google or Facebook). Note that certain traceroutes identified more than two distinct ASNs due to b-MNO's inter-AS routing patterns, involving different intermediary networks or global points of presence before reaching the destination service provider. 
We provide a more detailed analysis of observed ASNs in traceroutes in Section~\ref{sec:res:network:path:public}. 

Having ensured that each traceroute captures the CG-NAT within the PGW provider's core, we use the first public IP address as the demarcation point in our analysis; we label preceding hops as the \textit{private path} and subsequent hops as the \textit{public path}, differentiating between the initial routing within PGW provider's core and the routing after internet breakout.
We further support the path analysis with SGW locations inferred from volunteers' geolocations. This provided insight on GTP tunnels initially traversed by roaming packets, which traceroutes do not capture.   

Ultimately, our approach compiles a dataset for each traceroute, detailing path length, PGW provider, number of private and public hops, as well as IP address, geolocation, ASN, and Round-Trip Time (RTT) for each hop (when ICMP was supported).



\subsubsection{Private Path} Across the traceroutes, the latency difference between the first and last hop (which is the first hop with a public IP address) of the private path was generally negligible, with an average of 8.06 ms. This minimal difference in latency suggests a close physical proximity between the PGW and the CG-NAT. Consequently, we approximate the geolocation of the PGW by geolocating the first public IP address in the traceroute. We further refer to this address as the "PGW IP address" since it is assigned by the PGW provider.

Figure~\ref{fig:private_hops} shows the private path length, \ie the number of private hops, obtained from traceroutes to Google, distinguishing between country and network setup. Due to opaqueness of IPX-P network for end-to-end measurements, results for Facebook were equivalent and thus omitted. When the boxplots collapse to a single line, stable path lengths were measured, \eg in Pakistan, with 4 hops for SIM and 8 hops for eSIM. For visual clarity, each country on the x-axis is color-coded according to the architecture of its associated eSIM. 

We first compare private path lengths between native eSIMs (Korea and Thailand) and physical SIMs. 
In Thailand, no significant difference was observed: 
the 1,073 traceroutes from both SIM and eSIMs encountered 15 distinct PGW IP addresses (owned by AS9587, dtac) at a distance comprised between 4 and 10 hops. 
Conversely, the 252 traceroutes performed using physical SIM and eSIM in Korea encountered 35 and 16 unique PGW IP addresses, respectively, with no overlap. The traceroutes done via eSIM consistently had the private path length of 7 hops, with 16 PGW IP addresses all geolocalized in Seoul.  
Seoul also hosts 33 of the 35 PGW IP addresses accessed with the physical SIM, at a distance comprised between 7 and 9 hops. The two remaining PGWs are hosted in Goyang and Cheonan, at a distance of 7 to 8 hops.  

The above observation suggests that the physical SIM -- provisioned by U+ UMobile~\cite{uplusumobile2024}, an MVNO on top of LG UPlus -- may be subjected to a different routing policy compared to the eSIM. This observation loosely aligns with findings from~\cite{mvno_paths}, which indicate that MVNOs might route packets less efficiently than their parent MNO networks. 

For roaming eSIMs using HR (Pakistan and UAE), packets initially traverse from the v-MNO's network to the b-MNO (Singtel) via GTP tunneling. From their 1,803 traceroutes, we identify four PGW IP addresses within the IP range operated by Singtel (202.166.126.0/24), all geolocalized in Singapore. This suggests that IPX routing policy is consistent for HR traffic when the same base operator is involved.  

\begin{figure*}[t]
\centering
\begin{minipage}[t]{0.47\linewidth}
    \centering
    \includegraphics[width=0.9\textwidth]{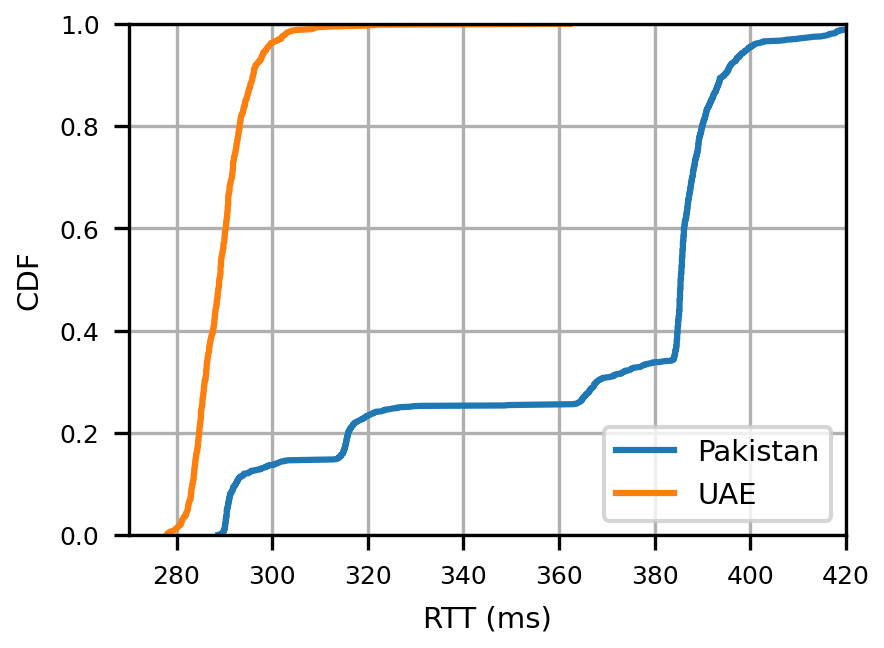}
    \vspace{-5mm}
    \caption{\small CDF of RTT towards Singtel PGWs from the HR eSIMs in Pakistan and UAE.}
    \label{fig:uae_pak_latency}
\end{minipage}
\hfill
\begin{minipage}[t]{0.47\linewidth}
    \centering
    \includegraphics[width=0.9\textwidth]{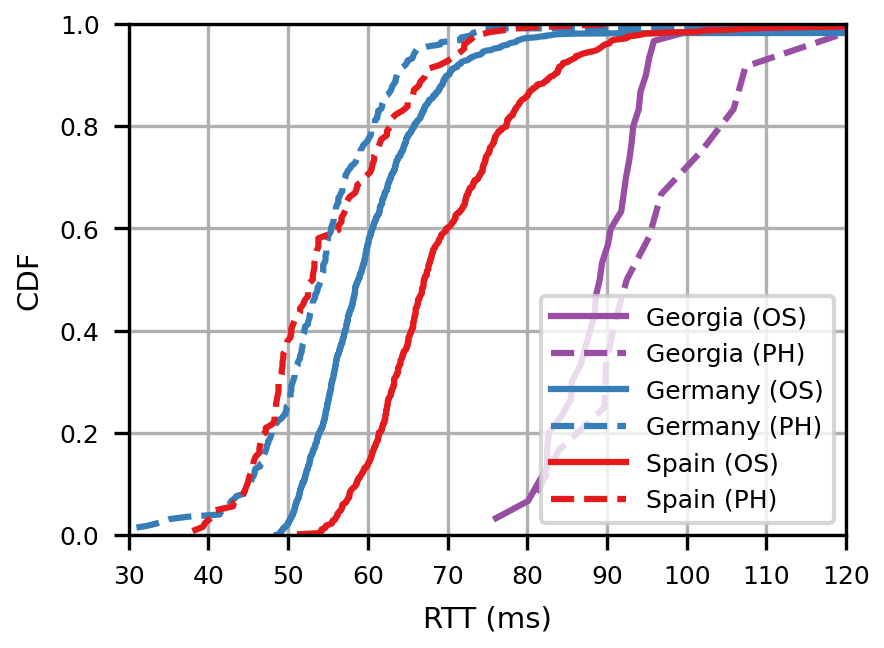}
    \vspace{-5mm}
    \caption{\small CDF of PGW RTT from IHBO eSIMs in Georgia, Germany and Spain. OS: OVH SAS; PH: Packet Host.}
    \label{fig:ovh_versus_ph}
\end{minipage}
\end{figure*}

To investigate the performance implications of the above observation, Figure~\ref{fig:uae_pak_latency} compares the Round Trip Time (RTT) towards Singtel PGWs from the two HR eSIMs deployed in Pakistan and UAE. The figure shows the Cumulative Distribution Function (CDF) of the ``best'' RTT values -- as indicated by traceroute -- at the hop where PGW IP was returned. We note that the RTT is shorter for the UAE eSIM, despite the same path length (4 hops) and being geographically farther from the PGW's location compared to the Pakistan eSIM. We conjecture that this difference could be the result of the v-MNO in the UAE (Etisalat) having better peering agreements with the IPX providers and/or the b-MNO (Singtel).


Finally, for roaming eSIMs using IHBO (green boxplots in Figure~\ref{fig:private_hops}), packets must traverse from the v-MNO's network to an IPX-P via GTP tunneling. The eSIMs for Georgia, Germany, UK, and Spain were provisioned by \textit{Play}, an MNO based in Poland. In Qatar and Saudi Arabia, they were provisioned by \textit{Telna Mobile}, an MNO based in the US. Despite the variations in their b-MNOs, traceroutes from IHBO eSIMs consistently resolve to one of the 2 PGW providers: OVH SAS (AS16276) and Packet Host (AS54825). We observe that most IHBO eSIMs alternate between these two PGW providers, except for the eSIM in Saudi Arabia, which relies exclusively on Packet Host. 

The two PGW providers differed in their internal routing behavior and public IP address assignment.  
For OVH SAS, six PGW IP addresses were identified, all reached within three hops. Most PGWs shared the same geolocation (Lille, France) except for one in Wattrelos (France). In addition, OVH SAS appears to assign PGWs for roaming traffic based on the b-MNO; Qatar eSIM (provisioned by \textit{Telna Mobile}) exclusively used one PGW IP address, while \textit{Play}-provisioned eSIMs alternated among the remaining 5. For Packet Host, we identified four PGW IP addresses, which are reached at either the 6th or 7th hop. This suggests potential load balancing within Packet Host's network core. Unlike OVH SAS, PGW IP addresses involving Packet Host were evenly distributed across different eSIMs, regardless of the b-MNO.

We now compare OVH SAS and Packet Host in terms of latency. Figure~\ref{fig:ovh_versus_ph} depicts the CDFs of RTT at PGW IP hops from eSIMs in Georgia, Germany, and Spain (all provisioned by \textit{Play}), with respect to the PGW providers. 
We exclude Qatar from this analysis due to the lack of statistically significant differences between the two PGW providers, likely due to the limited sample size. We also omitted Saudi Arabia because the eSIM exclusively utilized Packet Host.

The figure shows that in Germany and Spain, packets generally breakout to the internet faster when operated by Packet Host than by OVH SAS, despite the latter requiring half the number of hops. This trend was also observed in the UK, which we omit from the figure for brevity. This outcome contrasts with observations in Georgia, where Packet Host suffered from much higher RTTs especially in the fourth quartile. Statistical analysis did not support physical distance from the end-user as a factor influencing these latency differences (p > 0.05). Given the lack of visibility inside the IPX network, this result could be explained by several potential reasons: more efficient routing policies and/or load balancing strategies, or even the differences in the quality of interconnection agreements between the v-MNO and PGW providers.

\begin{figure}[t]
    \centering
    \includegraphics[width=0.9\columnwidth]{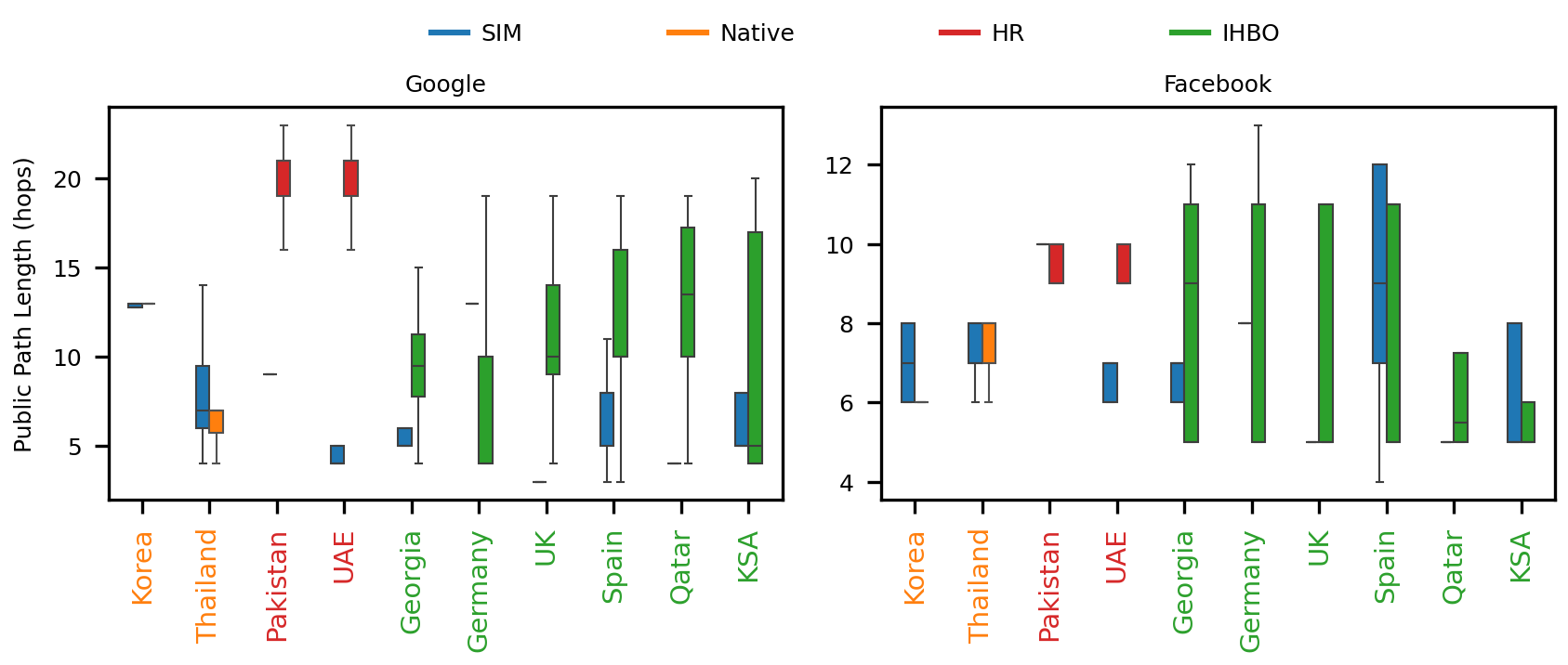}
    \vspace{-3mm}
    \caption{\small Public path length as a function of SIM/eSIM and country; traceroutes to Google (left) and Facebook (right).}
    \label{fig:public_hops}
\end{figure}


\subsubsection{Public Path}
\label{sec:res:network:path:public}
We now investigate the \textit{public} paths, \ie the hops recorded after packet's breakout to the internet. Figure ~\ref{fig:public_hops}  shows the public path length from traceroutes to Google and Facebook as a function of country and network configuration. As expected, the data paths measured from native eSIMs (Korea and Thailand) and corresponding SIMs are comparable.
Focusing on countries with roaming eSIMs, traceroutes to Google consistently required longer paths than physical SIMs. In contrast, the scenario for Facebook varied by country; traceroutes using physical SIMs in Pakistan, Spain, and Saudi Arabia, typically showed comparable or even longer paths compared to eSIMs. We conjecture that this stems from different peering arrangements between the SPs and PGW Providers~\cite{mvno_paths}. 
Across both SPs, traceroutes from roaming eSIMs exhibit higher variance in public path length compared to their SIM counterparts, as evidenced by the wide spread of box plots. 

Analysis of the AS path reveals similar inter-domain routing behavior across network configurations. Figure ~\ref{fig:unique_asn} plots the median number of unique ASNs observed in traceroutes to Google and Facebook, categorized by country and SIM configuration. Inter-domain routing was rare across most traceroutes, suggesting that the variability in the public hop counts (Figure ~\ref{fig:public_hops}) can be attributed to SPs' internal routing policies. Typically, traceroutes identified two unique ASNs that were indicative of direct peering between the PGW provider and the destination SP. However, a significant portion of traceroutes -- particularly those reaching Facebook via eSIM in Germany and both SIM configurations in Qatar -- revealed only the SP's ASN. This likely occurred due to the PGW provider's CG-NAT failing to respond within the traceroute timeout, possibly caused by router congestion or low-priority ICMP packet configuration.

In contrast, some countries involved more complex inter-AS routing patterns. 
Traceroutes in Spain via physical SIM consistently detected three unique ASNs including AS3352 (TELEFONICA DE ESPANA) and AS12956 (TELEFONICA GLOBAL SOLUTION), with the latter presumably serving as the global point of presence for the former. Similarly, traceroutes in Pakistan using the physical SIM (from Jazz) traversed IP addresses associated with AS23966 (LINKdotNET Telecom Limited) and its upstream carrier, AS38193 (Transworld Associates Pvt. Ltd.). Futhermore, eSIMs in Pakistan and the UAE, utilizing the same base operator (Singtel), traversed between three to five different ASNs, all geolocated within Singapore, before reaching their respective service providers. 

\begin{figure}
    \includegraphics[width=1\linewidth]{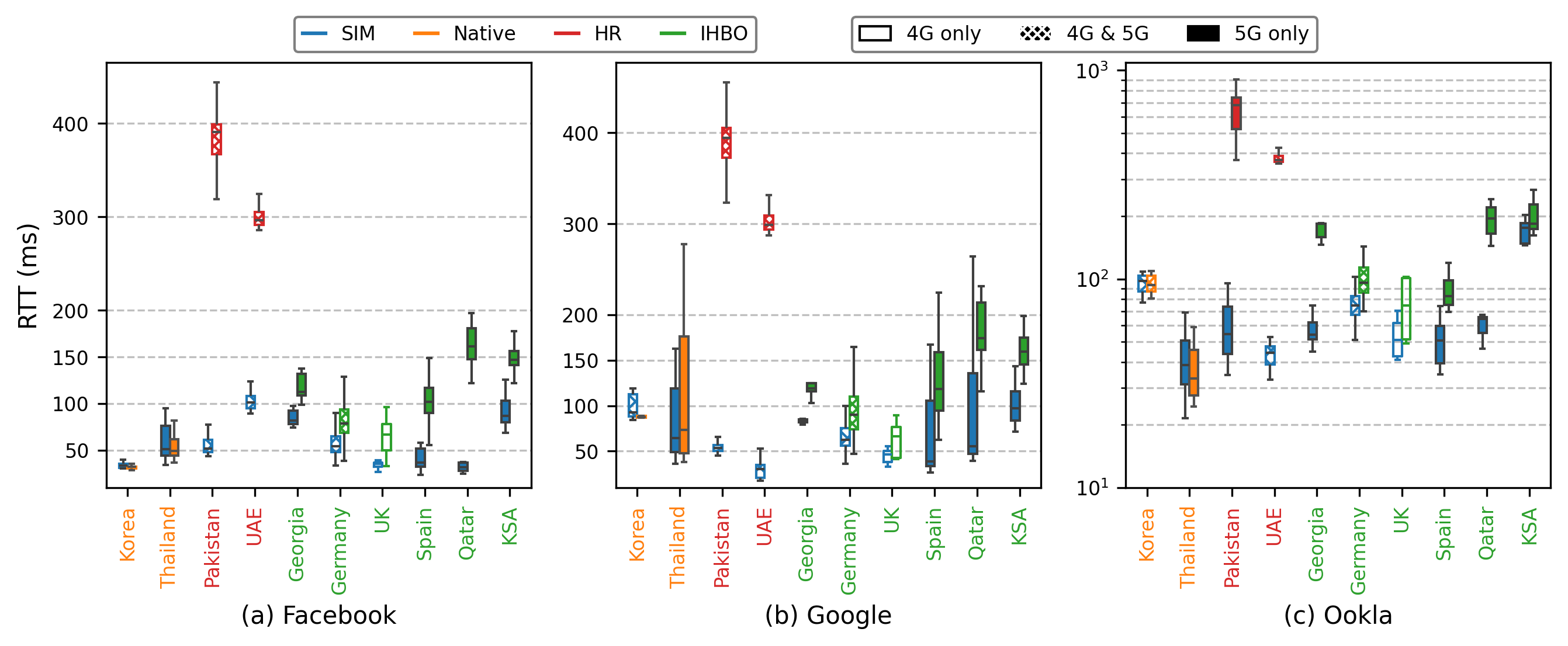}
    \vspace{-0.2in}
    \caption{\small 
    RTT measured in the last hop to traceroutes to (a) Facebook and (b) Google. (c) plots the latency to the closest Ookla Speedtest server from the PGW. The color and pattern of each boxplot indicates the corresponding network configuration and RAT(s) used during the measurement. 
    }
    \label{fig:low_level_ping}
\end{figure}
\section{Airalo Performance}
\label{sec:airalo_perf}
This section characterizes Airalo's performance. Specifically, we investigate: i) the performance differences between physical SIM and eSIM with the same v-MNO, and ii) the impact of the different network configurations adopted in Airalo. 


\vspace{-2mm}
\subsection{Network Performance }
\label{sec:performance}
\vspace{0.05in}
\noindent\textbf{Network Latency.} We measure network latency as the Round Trip Time (RTT) between mobile devices -- equipped with both physical SIMs and eSIMs -- and: i) two popular content providers (Facebook and Google), ii) Ookla, a widely used speedtest tool. We chose these service providers (SPs) for their global footprint, with edge servers strategically located close to most users. 

Figure~\ref{fig:low_level_ping} shows boxplots of latency to different SPs between physical SIMs and eSIMs across countries. We present latency data in boxplots grouped by country. Boxplots are further differentiated based on the Radio Access Technology (RAT) used during the measurements: empty boxplots refers to tests conducted over 4G/LTE, filled boxplots indicate tests conducted over 5G, and boxplots with a pattern comprise tests conducted via both 4G and 5G. The color coding in each boxplot reflects the network configuration; e.g., blue denotes a physical SIM from v-MNO.

Our analysis reveals similar latency patterns across SPs: in each country, roaming eSIMs tend to exhibit higher latencies compared to their SIM counterparts, with average increases of roughly 621\% and 64\% for HR and IHBO, respectively. 
Interestingly, despite the similar distances of approximately 6000~kms from their PGWs, IHBO eSIMs in Qatar and Saudi Arabia reported significantly lower latency compared to those by the HR eSIM in UAE. In fact, all RTT measurements involving HR eSIMs from Pakistan and UAE provided \textit{less desirable} latencies (\ie exceeded 150~ms) according to~\cite{hp2024}. We observe the most profound disparity in Pakistan while using 4G, where the eSIM was afflicted by a median RTT of 389 ms, versus 50 ms with the physical SIM. 

The figure also shows that for native eSIMs (orange boxplots for Korea and Thailand), such consistent performance degradation was not observed. We substantiated these observations through t-tests to assess the statistical relationship between RTTs and SIM configurations. 
In countries with roaming eSIMs, the p-value was 7.65368e-5, indicating that physical SIMs perform significantly better than eSIMs. In contrast, the p-value for countries with native eSIMs (Korea and Thailand) was 0.152, indicating no significant difference in latency between physical SIMs and eSIMs.

Figure~\ref{fig:low_level_ping} further shows higher variances in RTTs measured for roaming eSIMs, suggesting less consistent performance. We confirmed this through Levene's test~\cite{levene1960robust}, designed to evaluate the homogeneity of variances across groups. The resulting p-value of 0.025 confirms 
greater variability in RTTs for eSIMs compared to physical SIMs.

\begin{figure*}
    \centering
    \includegraphics[width=0.95\linewidth]{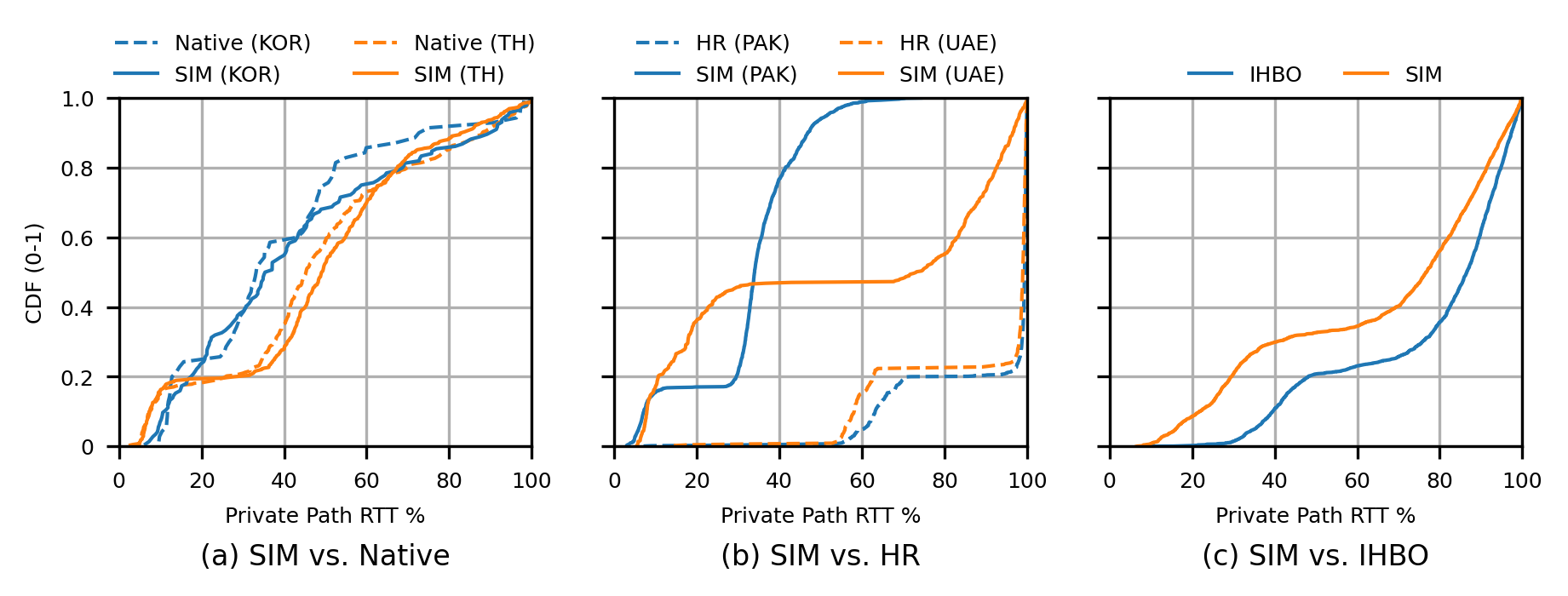}
    \vspace{-10pt}
    \caption{\small CDFs of the \% of latency which is \textit{private}, \ie due to RTT between mobile device and PGW. (a) compares SIM vs native eSIMs in Korea and Thailand; (b) compares SIM vs HR eSIMs in Pakistan and UAE; (c) data from 6 countries utilizing IHBO eSIMs.}
    \label{fig:cdf_pgw}
\end{figure*}

\vspace{0.05in}
\noindent\textbf{Private Path Latency.}  We augment the latency analysis by investigating the impact of \textit{private} paths, quantified as the percentage of RTT at the PGW hop -- where first public IP address was observed -- to the RTT at the final hop in traceroutes. 
Given that the private path differs vastly depending on the network configuration (see Section~\ref{sec:res:network:path}), we aggregate latency figures by the network setup associated with each eSIM, as shown in Figure~\ref{fig:cdf_pgw}. Note that physical SIMs consistently exhibit low private path latency, with mean of 31.06~ms and 95\% confidence interval of ± 0.78~ms. We use this stability to identify eventual disruption in the public internet; accordingly, the ``SIM'' curves in Figure~\ref{fig:cdf_pgw} captures the variability of the public internet latency in their respective regions. 

Figure~\ref{fig:cdf_pgw}-a shows negligible  differences in the impact of private latency on the total latency, when comparing SIM and native eSIMs (Korea and Thailand). The similar patterns in the CDFs for each country indicate that, overall, eSIM traffic is treated equally to SIM traffic when both are provisioned by the same MNO.  The slight variations in their private latencies reflect their differences in private path lengths (see Figure~\ref{fig:private_hops}). 

\begin{figure*}
    \centering
    \includegraphics[width=1\linewidth]{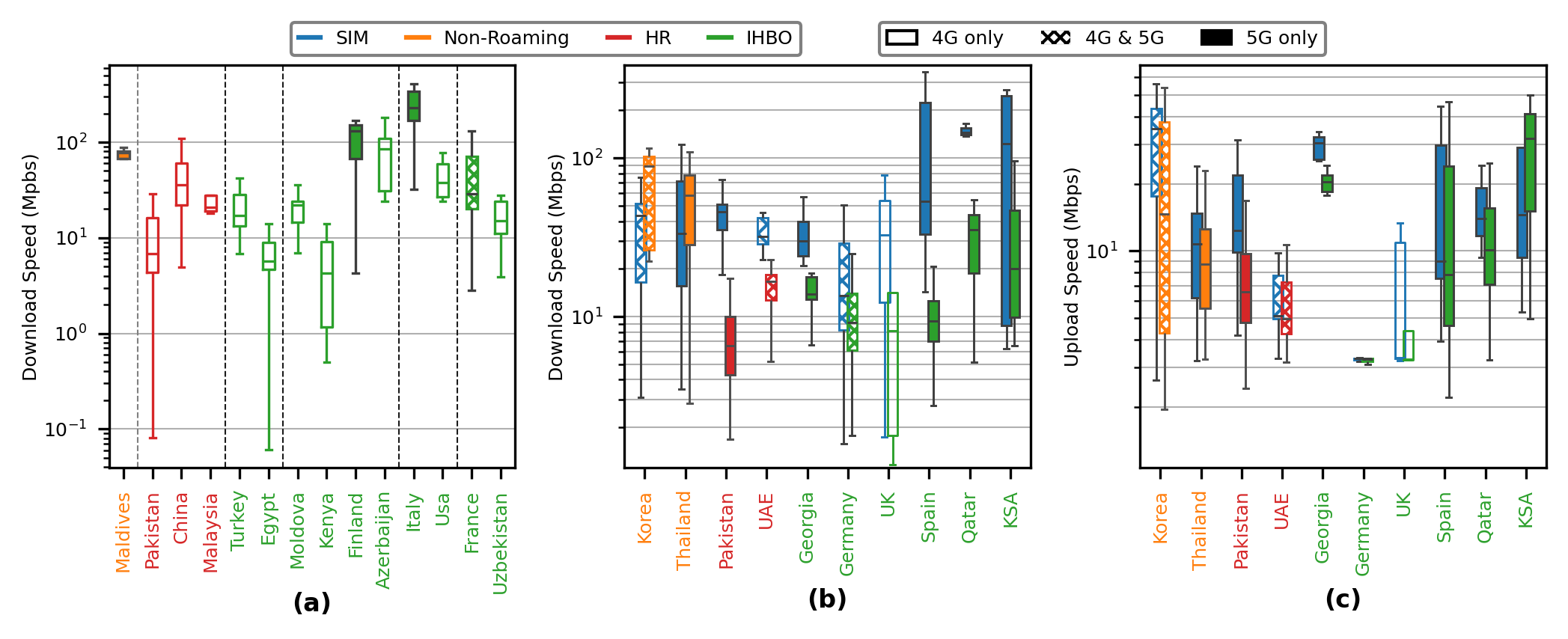}
    \vspace{-0.3in}
    \caption{\small (a) Download speed of eSIMs from web-based campaign. Dashed vertical lines group countries by network configuration and b-MNO. (b) Download and (c) upload speeds from device-based measurement campaign as a function of network configuration.}
    \label{fig:download_upload}
\end{figure*} 

In contrast, Figure~\ref{fig:cdf_pgw}-b shows a significant discrepancy between the CDFs of physical SIMs and HR eSIMs. For 80\% of traceroutes with the HR eSIMs in Pakistan and UAE, the private latency accounted for more than 98\% of the overall latency -- which instead only happens for less than 10\% of the measurements when using a physical SIM. Such inflation is the result of packets passing through the GTP tunnels between SGWs of the v-MNO and the PGWs of the b-MNO, which in this case was located in Singapore. 
This further underscores the implications posed by extensive GTP tunneling. 


Finally, Figure~\ref{fig:cdf_pgw}-c shows that IHBO could mitigate the latency associated with GTP tunnel traversals. Notably, the private latency was less than the public internet latency for 15\% of the measurements, compared to only 1\% of measurements for HR.
Furthermore, the CDFs for SIMs and eSIMs exhibit similar patterns, suggesting comparably low latency in the public internet. 
This low public latency is achieved by 
strategic placement of SP edge nodes near the PGWs, particularly in Western Europe (\eg France, Netherlands), where IHBO packets typically broke out. 
Our analysis implies that to further reduce the latency of roaming traffic to popular SPs, IPX network routing policies should aim to minimize GTP tunnel lengths by prioritizing the nearest available PGW, leveraging the global footprint of SP edge nodes, which are often close to most PGWs.

\vspace{0.05in}
\noindent\textbf{Download and Upload Speeds.} We  analyze Airalo's downlink using 117 \texttt{fast.com} measurements collected from 14 countries during the web-based campaign. Note that we do not control the devices from the web-based measurements, and as such, we lack visibility into additional information like channel quality. We summarize the results in Figure~\ref{fig:download_upload}-a, where vertical dashed lines group countries based on their eSIM network configuration and b-MNO, specifically eSIMs that utilize the same PGW providers. 
Our analysis reveals that eSIMs in countries geographically closer to their respective PGWs typically experience higher download speeds. For example, the median download speed in France (29~Mbps) was approximately twice as high as that in Uzbekistan (15~Mbps), despite both using PGWs located in Virginia (USA). However, there are exceptions to this trend. Notably, the eSIM in Azerbaijan experiences higher download speeds than the eSIM in Moldova, even though Moldova is closer to their shared PGWs in London. 

We now investigate how downlink (Figure~\ref{fig:download_upload}-b)  and uplink (Figure ~\ref{fig:download_upload}-c)  of Airalo eSIMs compared to physical SIMs, using Ookla speedtest performed during the device-based campaign. From a total of 749 measurements, we excluded those conducted under poor channel conditions (CQI < 7), resulting in 604 measurements (80\%). 
We use the results from native eSIMs (Korea and Thailand) to comment on whether eSIMs, by their nature, are subject to different network bandwidth than physical SIMs. In Thailand, both the SIM and eSIM exhibit similar download and upload speeds. In Korea, the eSIM generally achieved higher download speeds than the physical SIM, possibly because the latter was provisioned by a MVNO; past studies have shown that MVNOs could face traffic differentiation by their parent MNOs, potentially impacting bandwidth availability and user experience~\cite{oshiba2018accurate, mvno_paths}. 

In the remaining eight countries, physical SIMs showed higher download speeds compared to the roaming eSIM (either HR or IHBO). Visualized as red and green boxplots in Figure~\ref{fig:download_upload}-b: 78.8\% of roaming eSIM measurements fell into the slow download speed category -- as per the SpeedTest Global Index~\cite{globalIndex} -- (<= 15 Mpbs), whereas only 4.5\% reached fast download speeds (>= 30 Mbps). In contrast, for physical SIMs in the same countries, only 31.9\% of measurements were categorized as slow, while 48\% achieved fast download speeds. Additionally, there were large discrepancies in average download speeds among these physical SIMs, ranging from 13.6 Mbps in Germany to 137.2 Mbps in Saudi Arabia. 

A notable finding was that IHBO did not lead to significant improvement over HR in terms of download speed -- we observe comparable download speeds from the eSIM between UAE and Georiga, and between Pakistan and Spain. Furthermore, we observe a considerable variation among roaming eSIMs provisioned by the same b-MNO. For instance, the mean and 95\% CI of downlink (in Mbps) under 5G connection were 112 ± 2.16, 31.7 ± 2.26, 22.7 ± 1.98 in Spain, Georgia, and Germany, respectively. 

The eSIMs' upload speed was notably slower only in Pakistan and Georgia (p < 0.05), possibly due to stricter bandwidth policies by the local v-MNOs, as also observed for the download speed. For the remaining six roaming eSIMs, the upload speed wasn't as affected, indicating varying bandwidth policies for upload and download traffic among v-MNOs. This underscores how performance for roaming eSIMs depends heavily on v-MNO policies rather than roaming setup.

\begin{figure*}[t]
\centering
\begin{minipage}[t]{0.65\linewidth}
    \centering
    \includegraphics[width=1\textwidth]{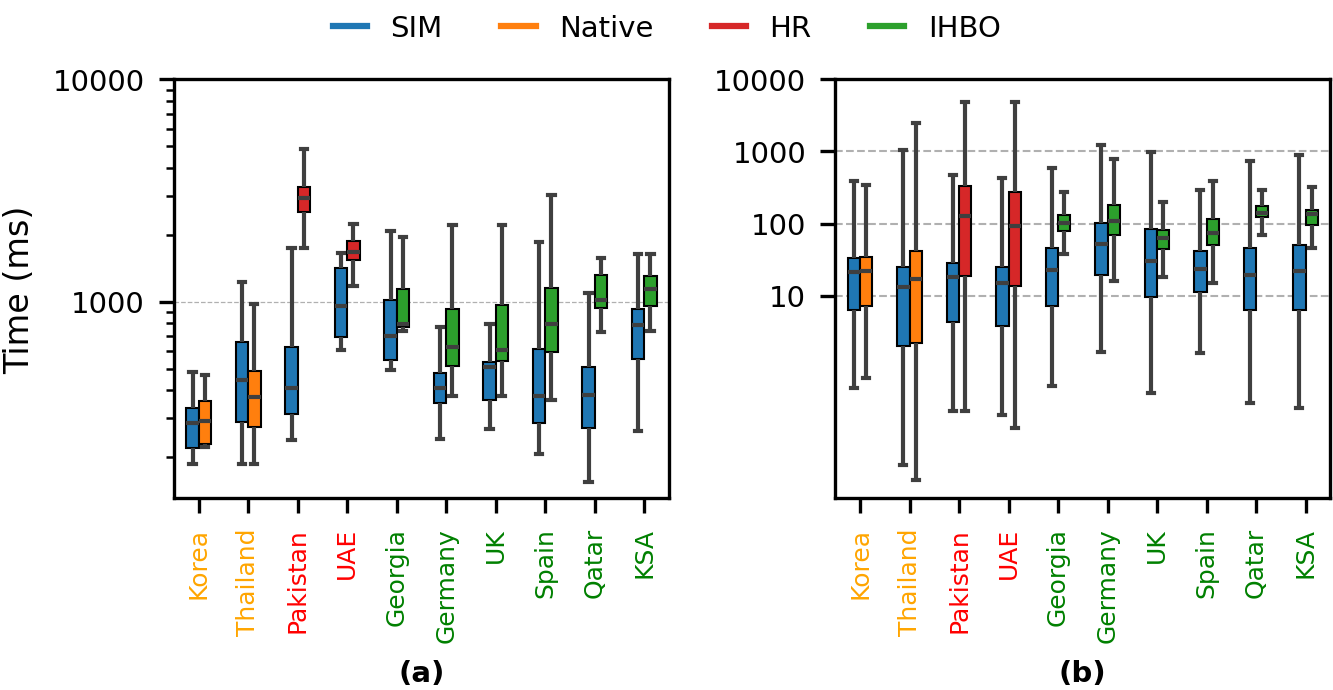}
    \vspace{-5mm}
    \caption{\small (a) Download time of the jQuery library via Cloudflare CDN. (b) DNS lookup time across countries as a function of network configurations}
    \label{fig:dns_cdn}
\end{minipage}
\hfill
\hfill
\begin{minipage}[t]{0.32\linewidth}
    \centering
    \includegraphics[width=1\textwidth]{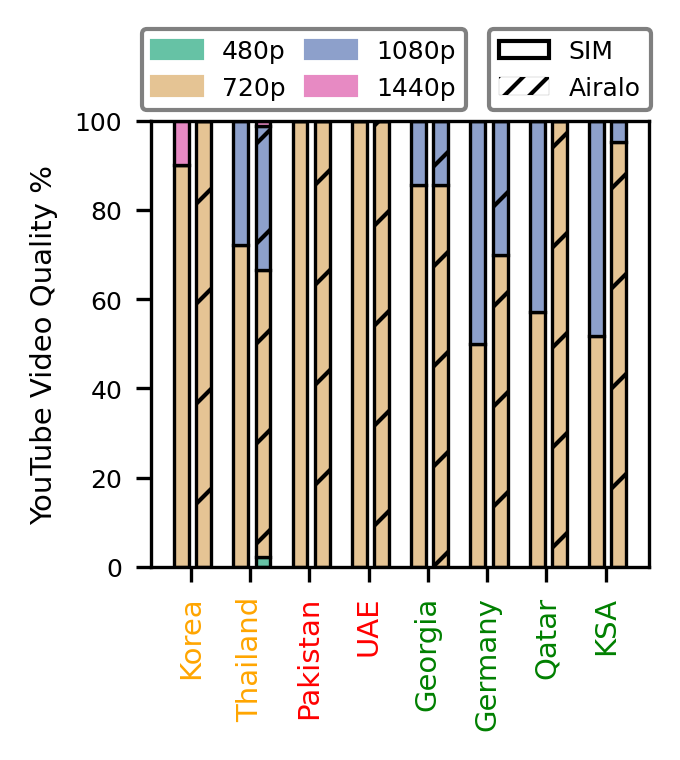}
    \vspace{-5mm}
    \caption{\small YouTube video res. across countries and network configurations.}
    \label{fig:youtube_quality}
\end{minipage}
\vspace{-0.12in}
\end{figure*}

\vspace{0.05in}
\noindent\textbf{CDN Download Time.} We measure the download time for ``jquery.min.js'' from six CDN providers via the device-based measurement. Given that all providers demonstrated similar trends across network configurations, we present the results from Cloudflare in Figure~\ref{fig:dns_cdn}-a and include results from the other providers in Appendix~\ref{sec:appendix:CDN}. Overall, the performance of native eSIMs and physical SIMs were comparable, paralleling observations on latency and bandwidth. Interestingly, the median download time from Cloudflare in Thailand was 18\% higher using the physical SIM, which can be partially attributed to a higher cache MISS rate of 7.7\%, compared to no cache misses with the eSIM.   
  
In contrast, significant performance degradation is evident for HR eSIMs in Pakistan and UAE (red boxplots in Figure~\ref{fig:dns_cdn}-a). On average, download times from all six CDNs were 523\% (Pakistan) and 320\% (UAE) slower over eSIMs than physical SIMs. In Pakistan, the average download time using the eSIM exceeded 3 seconds across all CDNs, indicating a substantial impact on web user experience.

IHBO eSIMs (green boxplots in Figure~\ref{fig:dns_cdn}-a) generally exhibit slower download speeds than physical SIMs, yet improve upon HR eSIMs. For Cloudflare, the average download time with IHBO eSIMs was 1,316 ms—higher than the two native eSIMs (306 ms in Korea and 514 ms in Thailand), yet clearly lower than that of the HR eSIMs (3,203 ms in Pakistan and 1,781 ms in the UAE). Across all CDNs, the increase in average download time for IHBO eSIMs over physical SIMs varied, ranging from 42.5\% in Germany to 174\% in Qatar.

\vspace{0.05in}
\noindent\textbf{DNS Lookup Time.} For network configurations where the PGW resides within the b-MNO's network (\ie physical SIMs, native eSIMs, and HR eSIMs), DNS resolution always occurs within the b-MNO. For example, in Pakistan, DNS queries from the physical SIM and HR eSIM are handled by their respective b-MNOs: PMCL (AS45669) and Singtel (AS45143). Conversely, IHBO eSIMS, which use PGWs external to the b-MNO's network, rely on Google DNS (anycast via 8.8.8.8 or 8.8.4.4). This approach aims to exploit Google's global footprint along with anycast routing to dynamically select DNS resolvers proximate to PGWs. 

To geolocate Google's DNS resolvers, we use Nextdns~\cite{nextdns}, an authoritative DNS with Time-To-Live set to zero, which ensures no DNS resolver should cache their responses. By forcing resolver misses, and adding a unique identifier to a user query, they identify the (unicast) IP address of a DNS resolver even in the presence of anycast. We then extract ASN and geolocation of these IP addresses via ipinfo.

For IHBO eSIMs, we find that 74\% of the DNS queries are directed to Google DNS resolvers located in the same country as the PGW. The greatest distance to a DNS resolver was observed for the US eSIM, involving a PGW in Dallas, Texas (operated by Webbing USA, see Table ~\ref{Tab:tab1}) and a DNS resolver in Tulsa, Oklahoma -- approximately 381~km apart. Occasionally, this eSIM was directed to a Google DNS resolver in Fort Worth, Texas, merely 20~km from the PGW. 

Our device-based measurement further explored the implications of network configurations for DNS performance. Figure ~\ref{fig:dns_cdn}-b plots the DNS lookup times across countries and network configurations. 
As expected, native eSIMs (Korea and Thailand) require short DNS lookup times, comparable to those of physical SIMs. On the contrary, substantial degradation in DNS lookup times is observed while using HR eSIMs (Pakistan and UAE). Compared to the physical SIM, the median DNS duration increases by 610\% and 517\% respectively, despite both PGWs and DNS resolvers being located within the same b-MNO (Singtel). This result again points to the limitation inherent to roaming technology, that extensive GTP tunnel traversals in the (private) IPX-network can significantly undermine any performance optimizations implemented on the public Internet.

The six IHBO eSIMs, which used Google DNS instead of b-MNO resolvers, also exhibited significant percentage increases in DNS lookup times over physical SIMs: ranging from 103\% in Germany to 616\% in Qatar. However, our analysis revealed that DNS over HTTPS (DoH) was employed for these eSIMs, and DoH is typically slower than unencrypted DNS due to the overhead of setting up TLS~\cite{bottger2019empirical}. DoH was used since this is a default setting in recent Android versions and we, unfortunately, \textit{forgot} to disable it. Note that other network configurations reverted to unencrypted DNS since MNO-operated DNS mostly do not support DoH~\cite{chhabra2021measuring, doan2021measuring}. 

\subsection{User Experience}

We conclude the section investigating the user experience of Airalo customers, specifically focusing on video streaming through YouTube. Figure~\ref{fig:youtube_quality} shows the distribution of video quality across countries from the device-based measurements, with the exception of Spain and the UK due to limited sample sizes. The figure shows that 720p is the most common resolution across countries and available network configurations.  The highest resolution observed was 1440p, recorded in 10\% of video playbacks on the physical SIM in Korea and 1.3\% on the eSIM in Thailand. However, the latter also reported the worst resolution, 480p, in 2.2\% of video playbacks.      

Results from IHBO eSIMs (green countries on the x-axis) show lower likelihood of 1080p streaming when using eSIMs compared to physical SIMs, with decreases of 20\%, 43\%, and 44\% in Germany, Qatar, and Saudi Arabia, respectively. This trend reflects the effect of lower download bandwidth allocated to roaming traffic (see Figure~\ref{fig:download_upload}). 

Georgia was an exception to the trend, with comparable streaming quality between physical SIM and the eSIM -- both configurations were equally likely to stream videos at 720p and 1080p. Similar patterns were observed in Pakistan and the UAE, where HR eSIMs maintained constant streaming quality at 720p, matching that of physical SIMs. This was a rather surprising finding given that the physical SIMs in these countries averaged download speeds of 7.9~Mbps and 8.3~Mbps, respectively, which are theoretically sufficient for higher-quality video streams (1080p, > 5~Mbps).  We conjecture that their b-MNOs may implement traffic differentiation, constraining bandwidth for YouTube, as suggested in~\cite{li2019deployed_traffic_diff, kakhki2015identifying}.

\begin{figure}[t]
\centering
\begin{minipage}[t]{0.45\linewidth}
    \centering
    \includegraphics[width=1\textwidth]{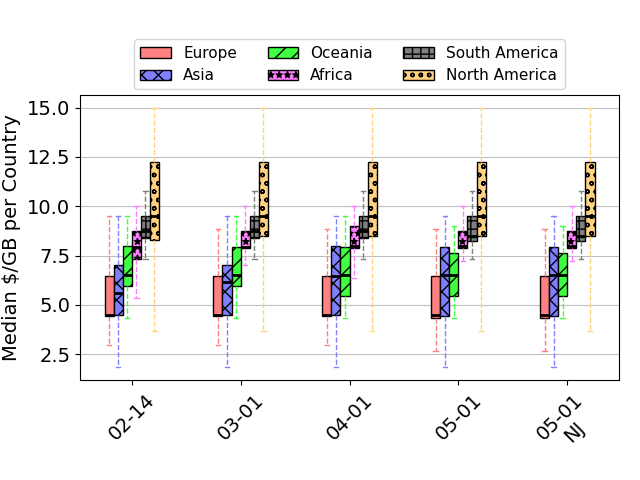}
    \vspace{-0.15in}
    \caption{\small Evolution over time -- Feb to May 2024 -- and space -- UAE and NJ -- of the \$/GB charged by Airalo. Each boxplot refers to the distribution of median \$/GB across countries in a given continent.}
    \label{fig:eco:airalo_time}
\end{minipage}
\hfill
\hfill
\begin{minipage}[t]{0.45\linewidth}
    \centering
    \includegraphics[width=1\textwidth]{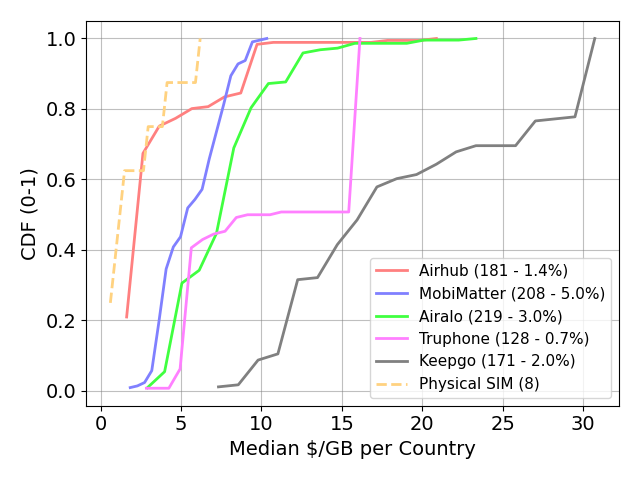}
    \vspace{-0.15in}
    \caption{\small CDF of median \$/GB per country (05/01/2024). The parenthesis holds the number of countries with at least 1 offer, and the \% of the 75,875 offers from EsimDB.}
    \label{fig:eco:comparison}
\end{minipage}
\vspace{-0.12in}
\end{figure}
\begin{figure}    
    \includegraphics[width=\linewidth]{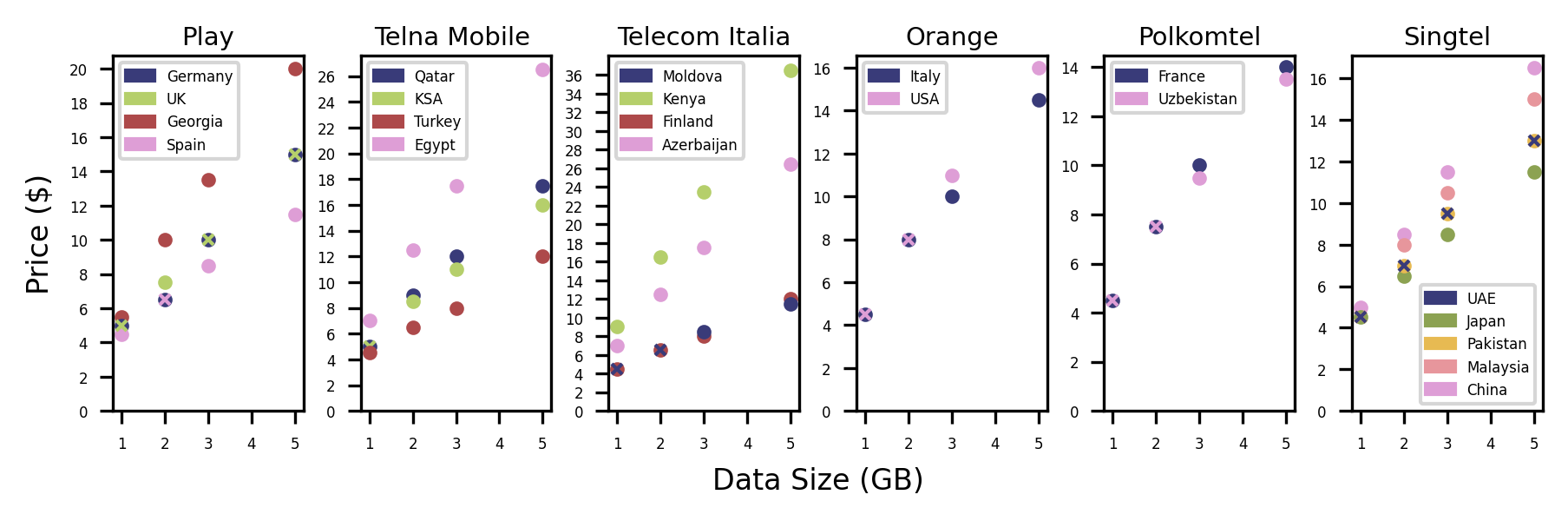}
    \vspace{-0.3in}
    \caption{\small Size (GB) versus price (\$) per eSIM and b-MNO.}
    \label{fig:hmno_price_plot}
\end{figure}

\section{Airalo Economics}
\label{sec:economics}
We conclude by investigating Airalo from an economic perspective. Given that Airalo offers thousands of plans, \eg 2,243 as of May 1st 2024 (9 plans per country, on average), Figure~\ref{fig:eco:airalo_time} shows the time evolution (February to May 2024) of the median cost per GB (\$/GB) across countries grouped by continent. The interested reader can refer to Figure~\ref{appendix:median_cost} in the Appendix for a visualization of the median cost per country. We further run our crawler at three locations (Madrid, Abu Dhabi, New Jersey) in April/May 2024, but only report one data-point from NJ, since no location impact was observed.

Overall, Figure~\ref{fig:eco:airalo_time} shows a significant price difference when comparing plans across continents; for example, the median cost per GB in Europe is about \$4.5, \ie half the price than in North America. It has to be noted that the main culprit of such high cost in North America stems from the plethora of expensive plans in Central America (see Figure~\ref{appendix:median_cost} in Appendix~\ref{sec:appendix:cost}). The figure also does not show dramatic cost changes over the last four months. There are, however, two changes to be reported. First, the median cost in Asia has increased from \$5.5 per GB (02-14-2024) up to \$6.5 per GB (04-01-2024). A similar increase is observed in Africa, with the 25th percentile growing from \$4.5 per GB up to \$6.5 per GB. Some minor fluctuations can be seen in other continents, but there is no major trend. Last but not least, no price discrimination was observed, \ie equivalent offers are presented to users from different regions, as suggested by the last boxplot which reports data crawled from New Jersey. 

Additionally, we compare Airalo's pricing with some of its competitors. Figure~\ref{fig:eco:comparison} shows the CDF of the median cost per GB across countries when considering a few interesting providers, which cover about 12\% of all the eSIM offers as per eSIMDB on 05/01/2024. The figure shows a median cost per GB ranging from \$2.3 at Airhub -- which has a presence in 181 countries -- and up to \$16.2 at Keepgo, which has a comparable footprint. MobiMatter has the majority of the offerings, \eg 5\% of the total versus 3\% from Airalo, but lacks coverage in a few countries; however, it is 60\% cheaper than Airalo, independent of the country. An interesting avenue for future work consists of leveraging our methodology to extend the analysis to other eSIM providers. 

The figure further shows the cost incurred by our volunteers to acquire a physical SIM card along with \textit{some} data. The figure shows that, overall, the cost per GB is the lowest when locally acquiring a physical SIM card. However, the \textit{total} cost incurred was overall higher than with Airalo, since most offers come with a larger data plan, \eg 40~GB in Spain for \$22.59, or require paying for a SIM, \eg \$15.72 in UAE.

Finally, Figure~\ref{fig:hmno_price_plot} compares the prices of various Airalo plans that share the b-MNO, further differentiating between plan sizes. For visibility reasons, we limit the plot to 5~GB, which covers about 70\% of the plans. Despite sharing the same b-MNO, the prices vary significantly across countries; for example, a Play Poland eSIM used in Georgia costs more than in Spain, up to twice as much as the data size increases. While this discrepancy likely stems from the distinct roaming agreements between b-MNO and v-MNO, the non-linear cost increase as a function of the data size seems unjustified. 

\section{Conclusion}
This paper presents the first measurement study of Airalo, a concrete example of a novel category of Mobile Network Aggregator (MNA) that we label as \textit{thick} MNA. The main innovation Airalo brings is a diverse eSIM marketplace, which combines native (sponsored) eSIM connectivity and roaming eSIMs, with their own deployment of internet breakout points in third-party infrastructure, decoupled from both the base and the visited Mobile Network Operators (b-MNO and v-MNO, in short). For example, we discover that Airalo offers eSIMs for Azerbaijan, Finland, Moldova, and Kenya, which use Telecom Italia as b-MNO. However, their roaming traffic consistently relies on IHBO (IPX Hub Breakout) by employing Packet Data Network Gateways (PGWs) in London, over infrastructure operated by Wireless Logic Limited.

With the rise of aggregators such as Airalo, the cellular ecosystem evolves in terms of complexity, with data paths that were once confined to a single operator realm now traversing multiple domains, and relying on resources from different entities (including the v-MNO, b-MNO, PGW provider, or IPX provider).
To map Airalo's underlying infrastructure, we ran a global campaign of web-based and device-based measurements covering 11\% of its footprint (24 of the 219 served countries). 
We find that most eSIMs (21 out of 24) are roaming (the other three are non-roaming native users of the b-MNO). Six b-MNOs provision these 21 eSIMs: 5 with Home-Routed Roaming (HR) via Singtel, a Singapore-based MNO, and 16 via IHBO. In theory, IHBO aims at optimizing roaming traffic by directing packets to a PGW located near the v-MNO. However, we find that, for 50\% of the IHBO eSIMs we measured, packets break out in PGWs that are farther away from the end user location than the b-MNO country. This stifles the potential of IHBO, despite overall improving upon the considerable latency impact of HR. 

The path fragmentation and the heterogeneity in terms of v-MNO radio capabilities make it difficult to identify a strong correlation between the underlying service configuration and the data rates at the end user. Overall, roaming network configurations did exhibit lower download bandwidth that native ones, which further impacted the user experience in terms of, for example, YouTube video resolution. 


Finally, we find that Airalo employs a consistent pricing strategy.
When compared to other eSIM providers, Airalo's offering is both \textit{large} -- second largest covering 219 countries -- and \textit{expensive} -- 11th position (out of 54 providers) with median cost per GB of \$7.9.  
An interesting avenue of future work is to expand our methodology to explore additional eSIM providers, to learn whether they also operate as thick MNAs, their architecture, and the level of performance they deliver to their users.

\bibliographystyle{ACM-Reference-Format} 
\bibliography{main}
\appendix

\begin{figure*}[!t]
    \includegraphics[width=1\linewidth]{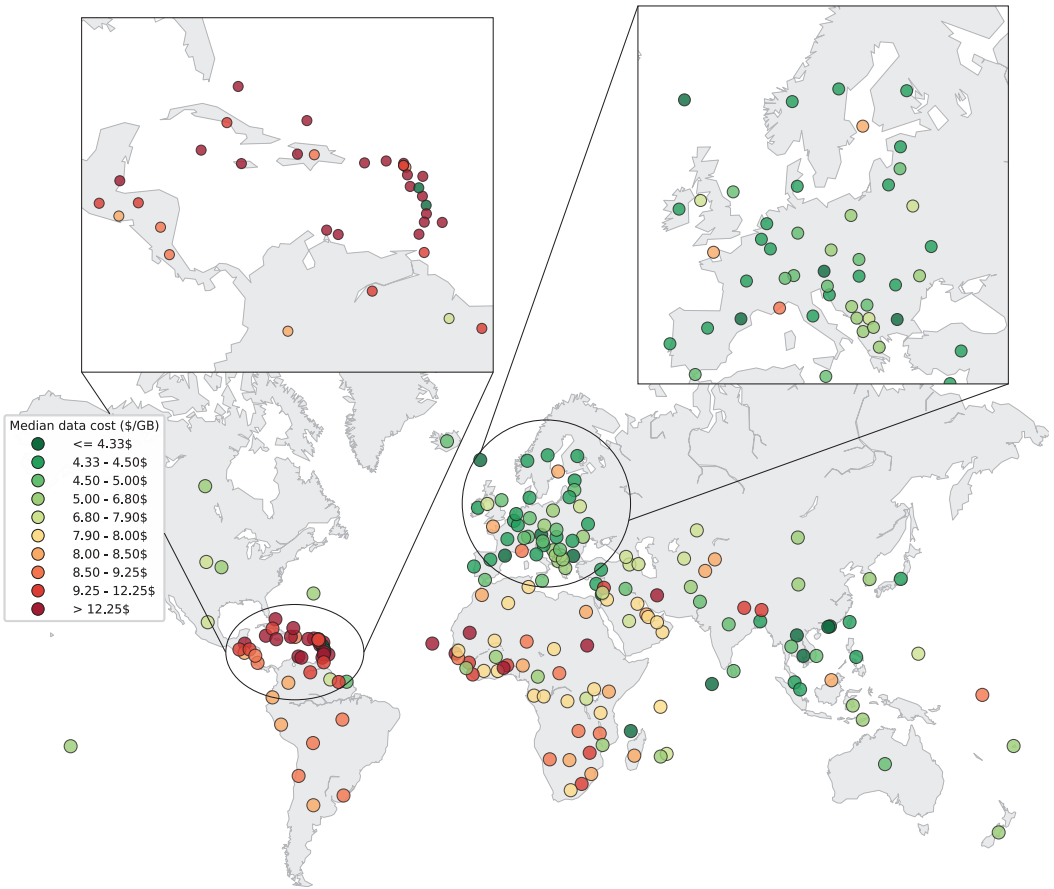}
    \vspace{-10pt}
    \caption{Median eSIM cost per country (\$/GB) from low (green, less than \$4.33) to high (red, more than \$12.25)}
    \label{appendix:median_cost}
\end{figure*}




\section{Airalo Cost Analysis per Country}
\label{sec:appendix:cost}
Figure~\ref{appendix:median_cost} visualizes the median eSIM cost per country, expressed as \$/GB. The location of each dot represents the country, while the color of each dot reflects the median eSIM cost in \$/GB. We divided eSIM costs into deciles, where each decile represents one tenth of the cost distribution (see Figure~\ref{fig:eco:comparison}), ranging from the lowest decile (dark green, $\leq$ \$4.33) to the highest decile (dark red, $>$ \$12.25). The figure provides more fine-grained details on the price difference observed per continent in Figure~\ref{fig:eco:airalo_time}. Further, it highlights that Central America (left circle in the figure) exhibits a consistent high cost per GB, \ie higher than the overall median cost worldwide (\$7.9) regardless of the country. 

\begin{figure*}
    \centering
    \includegraphics[width=1\linewidth]{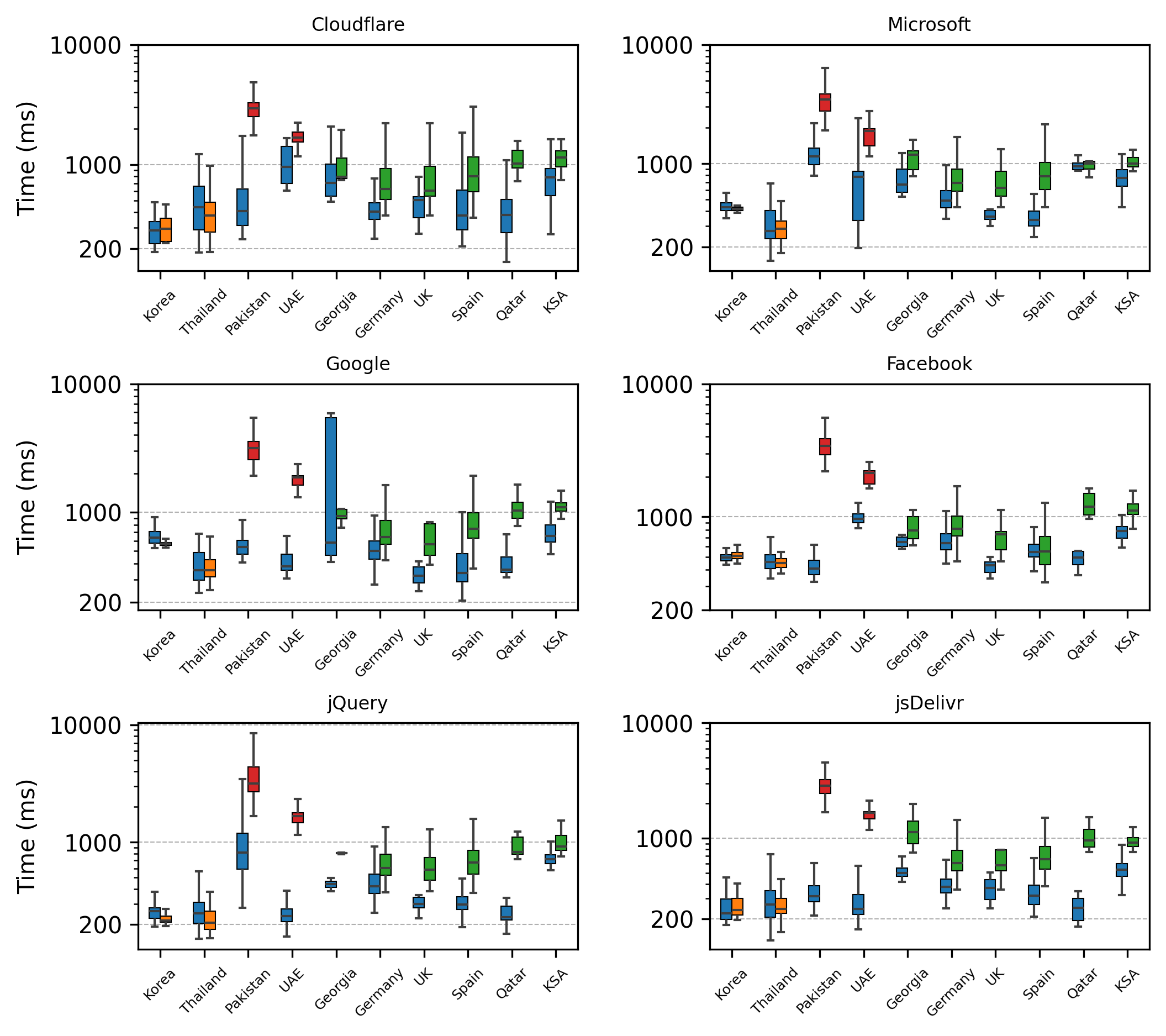}
    \caption{Time taken to download a jQuery library from six popular CDNS across countries and network configurations.}
    \label{fig:all_cdns}
\end{figure*}

\section{CDN Analysis}
\label{sec:appendix:CDN}
Figure~\ref{fig:all_cdns} compares the performance of six CDN providers across physical SIMs and eSIMs in our device-based campaign, measured as the download time for the last version of ``jquery.min.js``. 

Across all CDN providers, the download speeds exhibit a consistent pattern depending on the network configuration: native eSIMs typically achieve download speeds comparable to physical SIMs, while HR eSIMs experience notable performance degradation. This degradation is attributed to the technological limitations associated with data roaming. Although the six IHBO eSIMs also underperform relative to their physical SIM counterparts, the disparity in download speeds is less pronounced than that observed with HR eSIMs.

A notable exception was observed for Google CDN in Georgia, which occasionally experienced exceptionally slow download speeds, due to unusually long DNS lookup times that exceeded 5 seconds.

\section{Ethics}
\label{sec:appendix:ethics}
The underlying intention of our research was to assess the mobile network quality and performance of Airalo in multiple locations across the globe, measuring different metrics from low-level networking ones such as speed tests, DNS resolution, etc., to upper-layer application performance such as YouTube. Given that we recruited participants to carry custom-prepared mobile devices around, we obtained institutional review board (IRB) approval (IRB number is anonymized in order not to break double-blind) to conduct these studies. In addition, two of the authors have completed the required research ethics and compliance training, and are CITI~\cite{citi} certified. Participants were also provided with a consent form to read and sign, acknowledging their willingness to participate. They were given the opportunity to ask questions about the study and what was being collected. 

We asked the participants to carry these mobile phones, charge them, and install both a physical SIM and an eSIM we provided. We also instructed the participants not to use these phones or add any of their personal information or logins. As such, we do not collect any unidentifiable, sensitive, or personal information about the participants. The only foreseeable concern that might put our users' privacy at risk is the collection of the phones' GPS data, which in principle can reveal the participants movements. We did inform the participants before hand about this concern and we obtained their written consent that they approve this collection. As such, we believe that this are deemed to be of low-risk.

In our collaboration with the UK operator, the datasets we leverage in our research are protected under Non-Disclosure Agreements (NDAs) that explicitly forbid the dissemination of information to unauthorized parties and public repositories. The procedures for data collection and storage within the network’s infrastructure strictly follow the guidelines set forth by the MNO, and are in full compliance with local regulations. No personal and/or contract information was available for this study, and none of the authors of this paper participated in the extraction and/or encryption of the raw data. Ultimately, our datasets and research do not involve risks for mobile subscribers, while they provide new knowledge about the dynamics of virtual mobile operators.

\end{document}